\documentclass[lettersize,journal]{IEEEtran}
\usepackage{amsmath,amsfonts}
\usepackage{algorithm}
\usepackage{algpseudocode}
\usepackage{array}
\usepackage[caption=false,font=normalsize,labelfont=sf,textfont=sf]{subfig}
\usepackage{textcomp}
\usepackage{stfloats}
\usepackage{url}
\usepackage{verbatim}
\usepackage{graphicx}
\usepackage{cite}
\usepackage{xcolor}
\usepackage{amssymb}
\usepackage{subfig}
\begin{document}

\title{

Optimal In-Network Distribution of   Learning Functions for a Secure-by-Design Programmable Data Plane of Next-Generation Networks
}

\author{Mattia Giovanni Spina, Edoardo Scalzo, Floriano De Rango, Francesca Guerriero, Antonio Iera
        % <-this % stops a space
\thanks{The authors are with the University of Calabria, Italy}\thanks{M. G. Spina,  F. De Rango, and A. Iera are also with CNIT, Italy.}\thanks{This work was partially supported by the European Union under the Italian National Recovery and Resilience Plan (NRRP) of NextGenerationEU, partnership on ``Telecommunications of the Future” (PE00000001 - program ``RESTART”). This article has been submitted to Elsevier Computer Networks. }
}

% The paper headers
\markboth{Journal of \LaTeX\ Class Files,~Vol.~xx, No.~x, xxx~2024}%
{Shell \MakeLowercase{\textit{et al.}}: A Sample Article Using IEEEtran.cls for IEEE Journals}

\maketitle
\begin{abstract}
The rise of programmable data plane (PDP) and in-network computing (INC) paradigms paves the way for the development of network devices (switches, network interface cards, etc.) capable of performing advanced processing tasks. This allows running various types of algorithms, including machine learning, within the network itself to support user and network services. In particular, this paper delves into the deployment of in-network learning models with the aim of implementing fully distributed intrusion detection systems (IDS) or intrusion prevention systems (IPS). Specifically, a model is proposed for the optimal distribution of the IDS/IPS workload among data plane devices with the aim of ensuring complete network security without excessively burdening the normal operations of the devices. Furthermore, a meta-heuristic approach is proposed to reduce the long computation time required by the exact solution provided by the mathematical model and its performance is evaluated. The analysis conducted and the results obtained demonstrate the enormous potential of the proposed new approach for the creation of intelligent data planes that act effectively and autonomously as the first line of defense against cyber attacks, with minimal additional workload on the network devices involved.
\end{abstract}

\begin{IEEEkeywords}
In-Network Computing, Distributed AI, IDS, IPS Programmable Data Plane, Security by Design.
\end{IEEEkeywords}

\section{Introduction}
\IEEEPARstart{T}{he} evolving cyber threat landscape requires increasingly agile and adaptable cyber-security solutions. 

%The emerging paradigms of in-network computing (INC) and in-network distributed learning (INDS), coupled with the concept of distributed Intrusion Detection Systems (IDS), emerge as key components to address the challenge. 

In this context, a suitable integration of emerging paradigms, such as In-Network Computing (INC) and Distributed Intelligence, has the potential to revolutionize network security by offering robust, scalable, and resilient defense solutions against ever-evolving threats.

In-Network Computing exploits the idea of distributing computational tasks across the network infrastructure, rather than relying solely on edge or cloud computing resources. To this end, it leverages the capabilities of data plane network devices, such as switches, routers, and network interface cards (NICs), to perform data processing or caching. 
An interesting subfield of In-Network Computing that focuses on the use of distributed artificial intelligence (AI) techniques is the so-called “In-network distributed intelligence", which aims to enable network devices to collaborate and make intelligent decisions autonomously, without the need for centralized control.
This paradigm can make networks more scalable and fault-tolerant (as they become less dependent on centralized controls) and highly adaptable to changing conditions and traffic distributions in real-time through intelligent decisions about traffic routing, resource management, and network performance optimization.

Recently, interest is emerging in solutions that go beyond the standard uses of distributed intelligence on the network (such as supporting self-optimizing networks, Autonomous network management, and Context-aware networking), aiming to improve network security by allowing AI-enhanced network devices to autonomously distinguish between legitimate and anomalous traffic flows. This can, at the same time, improve the accuracy and increase the speed of intrusion detection.

As far as network security issues are concerned, the fixed-perimeter nature of traditional IDS and IPS solutions is no longer adequate for the highly pervasive and dynamic nature of next-generation networks.
Perimeter IDS and IPS in fact fail to be effective in case of attacks coming from inside the network that are continuously increasing. This is a consequence of the expansion of the attack surface resulting from the growing number of devices that will access next-generation networks in the future (IoT devices, vehicles, household appliances, etc.)

Even recent solutions, which rely on in-network telemetry and traffic data forwarding to a centralized SDN controller that runs the detection module and completes the decision-making process are not adequate to future networking scenarios.
%Next-generation networks require Active IDSs (also called Intrusion Prevention Systems - IPS), which leverage the INC and distributed intelligence paradigms to process and analyze network data 
These solutions may leverage sophisticated AI mechanisms to process and analyze network data in a centralized manner, with obvious inefficiencies in terms of increased control traffic and poor scalability. Alternative mechanisms leveraging distributed intelligence to process and analyze network data within Programmable Data Plane (PDP) devices, and enable the devices themselves to block threats through completely decentralized procedures can definitely improve the effectiveness and timeliness of intrusion detection, while ensuring greater scalability, resilience, and fault tolerance.

In this paper we refer to a new paradigm of Active Intrusion Detection Systems, we recently proposed in \cite{Spina2025}, which leverages the concept of \textit{AI model splitting} to split a Strong Learner (SL) model into its individual Weak Learner (WL) components. The latter are mapped into Virtual Network Functions (VNF), \textit{with both threat detection and response capabilities}, that can be distributed among the PDP devices of a next-generation network.

For the aforementioned paradigm to be truly effective, orchestration is required to always implement an \textit{optimal distribution} of learning functions that truly allows the network to experience
\textit{(i)} a good level of threat protection , 
\textit{(ii)} a reduced  processing load, and \textit{(iii)} a reduced impact on the standard functionality of the involved network devices (e.g., packet forwarding) and a reduced  reaction time to threats.

The main contributions of this paper can therefore be summarized as follows:
\begin{itemize}
%    \item demonstrate the potential of jointly using PDP devices and in-network distributed learning to enable the network user plane to implement a fully distributed active IDS, and increase the effectiveness of this new functionality;
    \item  propose an optimization model for efficient deployment of in-network learning models for distributed IPS, which balances security coverage with performance;
    \item propose a meta-heuristic approach providing a practical and scalable solution to the optimization problem;
    \item conduct a comprehensive performance analysis aimed at demonstrating the effectiveness of the proposed approach in enhancing the protection of the network against DDoS cyber threats while minimizing the impact on the overall network performance.
\end{itemize}

The remainder of the paper is organized as follows. Section~\ref{sec:back} presents the main related works in the key reference areas of this research. In Section~\ref{sec:prop}, the reference paradigm that exploits distributed in-network learning models to implement a “secure-by-design" data plane is introduced, while Section~\ref{sec:opt} illustrates a model for the optimization of the in-network distribution of learning elements and related meta-heuristic solution. The results of a comprehensive performance evaluation campaign are presented in Section~\ref{sec:perf}. Finally, in Section~\ref{sec:concl}, conclusions are drawn and future work is outlined.

\section{Related Works}
\label{sec:back}

%\subsection{In-Network Security: ML/DL-aided Traffic Classification}
%With the advent of Programmable Data Plane (PDP) and INC capabilities, recent efforts have focused on the design of in-network IDS solutions (also referred to as in-network classifiers) to address security-related challenges. 
Recent works in the literature address the issue of distributing computational functions relevant to AI (both training and inference), aiming to decompose a Deep Neural Network (DNN) into its layers to distribute the workload among a mobile Edge device and the Cloud. Among the optimization models proposed for this purpose, in \cite{neurosurgeon2017} the best split is determined by regression models predicting the computational and energy consumption of each DNN layer, while in \cite{autosplit2021} the optimal solution is determined considering the network and device resources usage to minimize the end-to-end latency between edge and cloud.

With the emergence of the potential of the in-network computing paradigm \cite{Kianpisheh2023Jan}, the focus has shifted towards a distribution of learning functions that also leverages the network segments connecting Edge and Cloud, namely Programmable Data Plane (PDP) and INC capabilities. Realizing the tight and crucial integration between AI and future 6G networks, the authors of \cite{NET4AIHuaweii}, \cite{schwarzmann2024new} and \cite{schwarzmann2023intelligent} have envisioned and analyzed the structural changes needed for future 6G networks to naturally accommodate distributed AI tasks within their Data Plane.

A significant area of research investigated the use of the programmable PISA (Protocol Independent Switch Architecture) switch architecture by means of Reconfigurable Match Tables (RMT), enabled by the introduction of the P4 language \cite{kim2016programming}. 
In \cite{siracusano2018network} the authors proposed N2Net, a solution that implements the forwarding pass of a Binary Neural Network (BNN) in a P4-enabled switch, outlining the limitations of modern programmable networking devices in accommodating complex ML/DL models characterized by intricate computations and mathematical operations. Following this direction, the authors of BaNaNa Split \cite{sanvito2018can} extended the use of the BNN to SmartNICs
%, proposing a solution 
to overcome the mentioned limitations: the joint work of programmable networking devices and end-host applications.
Nevertheless, the proposed solution does not fit well the concept of ubiquitous and pervasive in-network security, since it does not work without a server that shares the workload with the networking device. Instead, Saquetti et. al. \cite{saquetti2021} focus on the constrained nature of PDP devices as well as the limitations imposed by the reference PDP programming language (i.e., P4) when dealing with distributed intelligence in the network. Through a simple PoC – a neural network with 3 layers and a total of seven neurons – they proposed an optimization model to distribute the DNN within the network at single neuron granularity, with a one-to-one mapping between PDP and neuron. However, it turns out that this type of distribution is unfeasible when the NN is complex, severely limiting the applicability of the proposal.

With Taurus \cite{swamy2022taurus} and Homunculus \cite{swamy2023homunculus}, Swamy et al. proposed to equip the programmable networking devices with dedicated hardware capable of supporting map-reduce abstraction to perform complex mathematical operations. Main challenge of this approach is the need to redesign networking devices with custom and expensive hardware to enable them to perform ML/DL-relevant tasks.

Parallel efforts have focused on encoding ML models within programmable networking devices, particularly Random Forests (RFs) and Decision Trees (DTs). 
In this direction, SwitchTree \cite{lee2020switchtree} and Forest \cite{busse2019pforest} stand out as the most valuable examples. Both proposals strove to find the best encoding methodology to embed DTs and RFs within constrained and instruction set-limited PDPs.
Following this trend, the works in \cite{zheng2021planter,xie2022mousika,zheng2024iisy, xie2023empowering} show effort in designing a framework capable of encoding general RF/DT within P4-enabled networking devices. 
%Recent research has demonstrated the remarkable capabilities of eBPF (extended Berkeley Packet Filter), showing nearly equivalent performance to P4 in managing general-purpose tasks offloaded to networking devices \cite{gallego2024fast}. An important contribution in this domain is found in \cite{gallego2024machine}, where the authors focus on developing an efficient and effective encoding of a DNN using eBPF technology.

A common effort emerging from the literature is the search for optimal encodings of the entire (sometimes complex) ML/DL models to adapt them to network devices with reduced impact on packet forwarding performance. None of them addresses how to intelligently distribute in-network classification modules to achieve pervasive and ubiquitous security through a fully distributed and collaborative approach of such modules, which is the objective of the novel paradigm studied in our paper. The potential of the approach described is accompanied by new challenges, such as finding the optimal positioning within the network of the AI-empowered security capabilities mentioned above to minimize both the delay in completing tasks and the resource consumption of the network devices involved. 
Our paper aims precisely to contribute to finding a solution to this compelling research problem.

\section{ Deployment of an ML-enabled IPS in a network dataplane}
\label{sec:prop}

The reference framework for the research reported in this paper is the one presented by the authors in \cite{Spina2024Oct}, where a new paradigm, according to which anomaly detection capabilities are deployed in a fully distributed way within devices of a typical data plane of a future programmable network, is introduced. Only the proposed innovative architecture and a simple proof-of-concept study were the subject of that paper in which among the most important future issues highlighted and left open there is the design of an optimization policy for the distribution of the learning functions within the most suitable data plane devices and their related chaining. The latter key issue is precisely the one addressed in the present paper.

\subsection{Projecting the Ensemble Learning over the Network}
For the benefit of the reader, we briefly report a few basic concepts, referring to the aforementioned article for the details of the hypothesized architecture. 

The reference framework includes all the functionalities to implement the proposed paradigm, distributed over three logical levels, \textbf{A}rtificial \textbf{I}ntelligence \textbf{P}lane (\textit{AIP}), \textbf{C}ontrol \& \textbf{O}rchestration \textbf{P}lane (\textit{C\&OP}), \textbf{AI}-\textbf{e}nhanced \textbf{P}rogrammable \textbf{D}ata \textbf{P}lane (\textit{AIePDP})
\cite{Spina2024Oct}.

\begin{figure}[!h]
\centering
\includegraphics[width=3in]{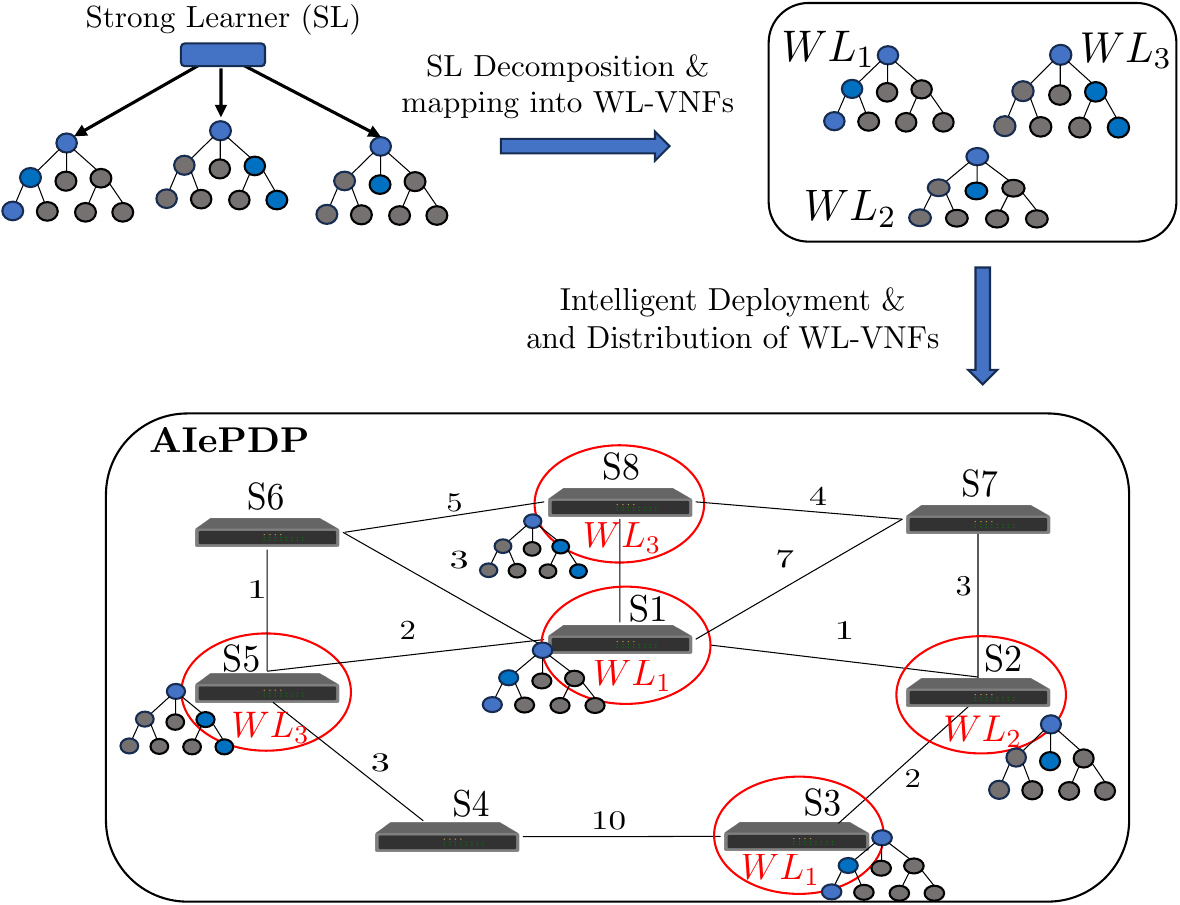}
\caption{Proposed Split-AI In-Network Distribution Strategy.}
\label{fig:split}
\end{figure}

It is envisaged that through ad-hoc functions included in the first level, the ML model to be distributed in the PDP is trained and then properly broken down into individual WLs coded as WL-VNFs that will cooperate with one another to perform attack detection and response. These WL-VNFs are made available to the orchestration functions that are in the second level. Here, an optimal distribution strategy of the WL-VNFs among PDP devices is decided, which allows the selected switches that host the WL-VNFs to operate cooperatively as an active IDS in the network.
The whole process is depicted in Fig.\ref{fig:split}.

The activities described in this paper refer to what was only theorized at the second level of the mentioned architecture, but not previously developed. Specifically, the goal is to find the set of WL-VNFs and the switches that shall host them in such a way as to maximize the security coverage of the considered network, i.e., the effectiveness in detecting and reacting to the maximum number of attacks.

The functions that will then perform this activity are hosted in the lowest level of the architecture (as shown in Fig.\ref{fig:split}), i.e. the AI-enhanced programmable data plane. Here, the cooperative policy that the group of WLs implements provides that all flows are analyzed and the suspicious ones are properly marked by each WL to  signal this to the following WLs that must be executed on the flow to reconstruct the original SL.

The flow, as it passes through the network, is analyzed by the various WLs that constitute the SL, and each one signals the result of its inference to the others. If a WL realizes that it is the last of the set that constitutes an SL and that all the others have already performed the flow analysis, it completes its analysis, and through a majority voting algorithm takes the final decision, blocking the flows that the WLs chain deems malicious. The algorithm is \textit{completely distributed} and does not require human involvement or intervention of the network controller. 
To allow distributed WL-VNFs to inform each other on the inferences carried out for a network flow, a custom header, P4-encoded, is considered as well as a procedure carried out by the PDP device augmented with the WL-VNFs.

\section{Problem Formulation}\label{sec:opt}
In this section, we propose a variant of the shortest path problem to optimize the deployment of the WL-VNFs.

Without losing generality, we will refer to a specific type of vulnerability of the programmable network considered, namely a DDoS attack in which the attacking hosts could also be compromised devices operating within the network. Therefore, it is assumed that the attacks can come not only from outside the network and therefore counteracted with a perimeter defense system (traditional, such as a firewall, or even innovative such as a entire AI model embedded in the entry/exit node of the network) but also from hosts inside the network that is intended to be protected. 
The specific optimization does not depend on the type of attacks -- that is instead related to the source of data used to train the SL and its WLs. The obtained results in terms of efficiency and effectiveness can, therefore, be generalized to other type of attacks provided that the SL is trained on traffic profile patterns that encode the attacks that one desires to detect. 
%Obviously, the specific optimization will depend on this type of vulnerability chosen, but the results obtained in terms of efficiency and effectiveness can also be generalized to other types of vulnerabilities provided that the type of WL-VNFs used is changed (i.e., they must be trained to detect and block threats of different nature).

\subsection{SL model decomposition into component WLs}

As previously mentioned, the proposed split-AI technique resorts to the Ensemble Learning theory. More specifically, the core idea of the proposal lies in the disaggregation of complex Ensemble Strong Learner (SL) models. 
In the context of ensemble learning, a \textbf{Strong Learner} is a model that achieves high predictive performance by \textit{combining the outputs of multiple individual weaker learners}, i.e., models that perform only slightly better than random guessing (they achieve an accuracy at least greater than 50\%).
Formally, let $\mathcal{SL} = \{WL_1, WL_2, \dots, WL_T\}$ be a set of weak learners where each $WL_i : \mathcal{X} \rightarrow \mathcal{Y}$ is a model trained on the input space $\mathcal{X}$ and output space $\mathcal{Y}$. 
In addition, given $\mathcal{I}=\{f_1, f_2, ..., f_n\}$ the set of input features on which the $\mathcal{SL}$ has been trained. Each $WL_i \in \mathcal{SL}$ is trained considering a subset $I_i\subset \mathcal{I}$ of input features, such that:

\begin{equation}
    WL_i\ \text{trained on}\ I_i\subset \mathcal{I}\ \forall\ WL_i\ \in \mathcal{SL}, \text{with}
\end{equation}
\begin{equation*}
    \bigcap_{i=1}^{T} I_i \ne \emptyset
\end{equation*}

%Additionally, according to Ensemble Learning theory, a model qualifies as a \textit{Weak Learner} only if it performs slightly better than random guessing, i.e., it achieves an accuracy at least greater than 50\%.

Given the mentioned formulation, a strong learner $SL : \mathcal{X} \rightarrow \mathcal{Y}$ is defined as an aggregation of the weak learners, such that:

\begin{equation}
    \mathcal{SL}(x) = \mathcal{A}(WL_1(x), WL_2(x), \dots, WL_T(x))
\end{equation}

where $\mathcal{A}(\cdot)$ is an aggregation function (e.g., majority voting for classification, weighted average for regression) that combines the independent intermediate inferences computed by each single WL in the SL. The number $T$ of WLs that compose the $\mathcal{SL}$ is required to be odd, to ensuring a majority-based aggregation function can always reach a unique decision. It is worth noticing that the SL is constructed through a bagging-based training process, which ensures that the resulting WLs are mutually independent. 

The described AI technique naturally aligns with networking principles, particularly resembling the concept of VNF chaining.

In the proposed \textit{Projecting the Ensemble Learning over the Network} split-AI technique, the $\mathcal{SL}$ is disaggregated and its weak learners are distributed and deployed onto programmable network elements. Let $\mathcal{P} = \{N_1, N_2, \dots, N_t\}$ be the ordered set of programmable network nodes traversed by a data flow. Each node $N_i$ hosts the weak learner $WL_i$ -- encoded as VNF --  such that:

\begin{equation}
    N_i \mapsto WL_i(x_{I_i}) \quad \text{with} \quad x_{I_i} = \pi_{I_i}(x)
\end{equation}

where $\pi_{I_i}$ is the projection operator selecting the subset $I_i$ of features from $x$.

The inference result is computed incrementally by each involved device in a distributed fashion. The final decision is produced by the last traversed node $N_t$ of the path, which applies the aggregation function:

\begin{equation}
    \mathcal{SL}(x) = A\left( \{WL_i(x_{I_i})\}_{i=1}^{t} \right) = A\left( r_1, r_2, \dots, r_t \right)
\end{equation}
computed at  $N_t$. This is achieved through a custom packet header that carries the intermediate results computed by each WL-VNF, allowing subsequent components to access these partial outputs and ultimately reconstruct the final result of the original SL. %\textcolor{red}{Metto l'immagine dell'header e lo spiego, o lascio cosi?****direi che è inutile...evitiamo...}
Notably, the WLs composing the SL are mutually independent, and as such, no ordering constraints are required during their traversal. This means that once a node $N_t$ has all the required intermediate inferences, it compute the $A(\cdot)$ to aggregate them and produce the final outcome.

\subsection{Optimization problem definition}
\begin{figure}[!h]
\centering
\includegraphics[width=3in]{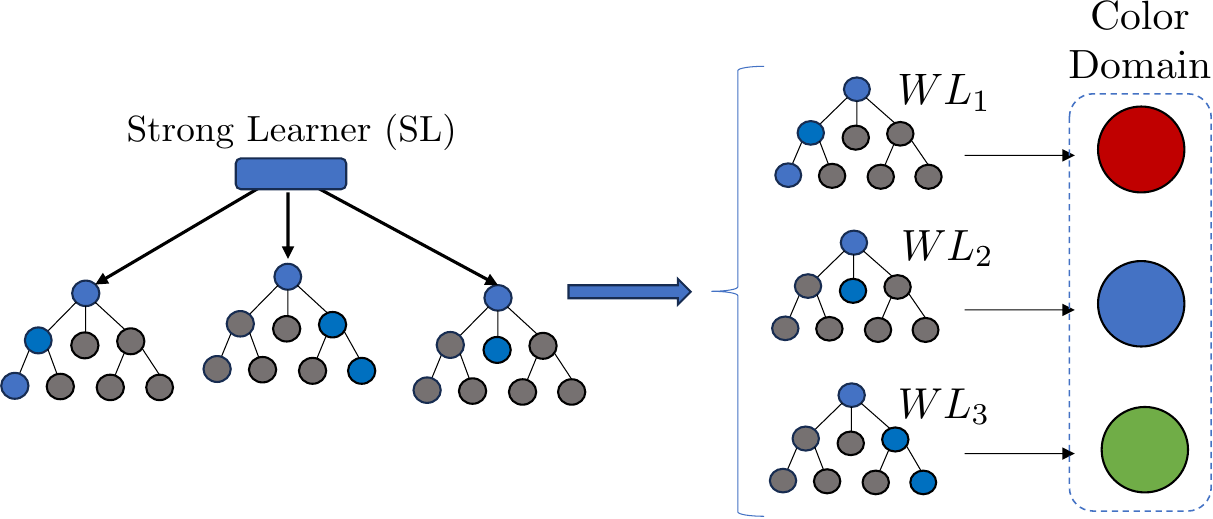}
\caption{From WL-VNFs to Colors domain.}
\label{fig:color_domain}
\end{figure}

Given the described Split-AI technique, once a complex SL is disaggregated into its lighter individual but functional components, an optimal deployment strategy is needed in order to place them on programmable networking devices accounting for a trade-off between network security and effective network management. 
In this direction, we represent a network by using a graph. The nodes in our model represent the network nodes in which the WL-VNFs can be deployed, while the edges denote the
links between network elements. We use node coloring to represent the implementation of specific WL-VNFs, where each color\footnote{The terms “color" and “SL/WL-VNF" will be used interchangeably. More specifically, SL-VNF refers to a scenario in which only one color is needed.} corresponds to a different type of WL-VNF and the coloring cost corresponds to the associated implementation cost. For instance, an SL composed of three WLs will determine three WL-VNFs and therefore three different colors (e.g., red, green, and blue), as shown in Fig.\ref{fig:color_domain}.

The graph edges are weighted to reflect a network connection characteristic, such as latency or bandwidth.
Our objective is to find the optimal deployment of WL-VNFs to ensure comprehensive network security coverage.

This approach guarantees pervasive and ubiquitous network protection, aligning with the need for robust cybersecurity measures in the evolving landscape of next-generation networks. Practically, we modified the behavior of the shortest path problem by adding and taking into account coloring constraints designing and introducing a new model named All-Pairs Shortest Path Coloring problem (APSPC), where the cost to be minimized includes both the costs of the different paths between pairs of source nodes and target nodes, ensuring that each path passes through at least one colored node for each color and the cost of coloring the nodes themselves.
In the remainder of the section, we propose a detailed mathematical model that represents the problem 
and a meta-heuristic approach, based on a Biased Random-Key Genetic Algorithm (BRKGA), providing a practical and scalable solution to the optimization problem.
\subsection{Exact Model}\label{sec:exact_model}
This section delves into the mathematical complexities of the APSPC problem through the development of an Integer Linear Programming (ILP) model. The problem is formulated on an undirected connected loopless graph $G=(V, E)$, with the goal of determining the simple shortest paths between all pairs of nodes (source-target) such that each path includes at least one vertex colored for each color in the set $C$. Despite the undirected nature of the graph, this model incorporates directed flow constraints, which are necessary for the formal definition of paths from a source node \textit{s} to a target node \textit{t}.
For this reason, with the abuse of terminology, once the nodes \textit{s} and \textit{t} have been fixed, any node can have outgoing and incoming edges.
Three sets of binary variables are introduced to indicate whether an edge is traversed and whether a vertex is colored with a specific color; specifically, let $x^{st}_{ij}$ be a binary variable equal to 1 if and only if the edge $(i,j)$ is visited in the path $s$--$t$, and $y_{ic}$ be a binary variable equal to 1 if and only if the vertex $i$ is colored by \textit{c} in the graph.
The last set of variables keeps track of the coloring of the nodes in each path \textit{s}--\textit{t}. In particular, given the color \textit{c}, fixed the source \textit{s} and the target \textit{t}, $z_{ic}^{st}$ must be equal to 1 if and only if in the path \textit{s}--\textit{t} the vertex \textit{i} is colored with \textit{c} and is traversed.
In addition, let $w_{ij} \in \mathbb{Z}^+$ be the positive weight associated with each edge $(i,j)$ and $p_c \in \mathbb{Z}^+$ the cost of coloring a node with color \textit{c}.

The All-Pairs Shortest Path Coloring problem presented can be formulated using the following programming model.

\begin{scriptsize}
\begin{eqnarray}
&&\min\sum_{(s,t)\in V\times V}\sum_{\substack{(i,j)\in E:\\ i\neq t \wedge j\neq s}} w_{ij}\cdot x^{st}_{ij}+\sum_{(i,c)\in V\times C}p_c\cdot y_{ic}\label{eq:ob_function}\\
&&s.t.\nonumber\\
&&\sum_{j\in V\setminus\{s\}}x^{st}_{ij}-\sum_{j\in V\setminus\{t\}}x^{st}_{ji}=
                \begin{cases}
    			1 &\text{if $i=s$}\\
    			-1 &\text{if $i=t$}\\
    			0 &\text{otherwise}\\
    		    \end{cases}\hspace{0.5cm}  \forall \; i,s,t \in V \label{eq:flow_conservation}\\
&&\sum_{(i,j)\in E(S)}x_{ij}^{st} \leq \sum_{i\in S\setminus\{k\}}\sum_{j\in V\setminus\{s\}}x_{ij}^{st} \hspace{0.5cm}   \substack{\forall s,t\in V;\forall k\in S;\\\forall S\subsetneq V\setminus\{s,t\}:|S|\geq 2} \label{eq:subtour_elimination}\\
&&\sum_{c\in C} y_{ic} \leq 1 \hspace{1cm}  \forall  i\in V\label{eq:coloring_1}\\
&&\sum_{i\in V}z_{ic}^{st}\geq 1 \hspace{1cm}  \forall \; s,t \in V;\forall c\in C\label{eq:coloring_2}\\
&& z_{ic}^{st} \leq \sum_{j\in V\setminus\{s\}}x_{ij}^{st}\hspace{1cm}  \forall \; s,t\in V;\forall i\in V\setminus\{t\};\forall c\in C\label{eq:coloring_3}\\
&& z_{tc}^{st} \leq \sum_{j\in V\setminus\{t\}} x_{jt}^{st}\hspace{1cm}  \forall \; s,t \in V;\forall c\in C\label{eq:coloring_4}\\
&&z_{ic}^{st}\leq y_{ic} \hspace{1cm}  \forall \; s,t,i \in V;\forall c\in C\label{eq:coloring_5}\\
&& x^{st}_{ij}\in\{0,1\}\hspace{1cm}  \forall s,t,\in V;\forall (i,j)\in E\label{eq:domain_1}\\
&& y_{ic}\in\{0,1\}\hspace{1cm}  \forall  (i,c)\in V\times C\label{eq:domain_2}\\
&& z^{st}_{ic}\in\{0,1\}\hspace{1cm}  \forall  s,t,i\in V;\forall c\in C.\label{eq:domain_3}
\end{eqnarray}
\end{scriptsize}

The objective of the model (\ref{eq:ob_function}) is to minimize the total weight of the traversed edges and the cost of coloring the nodes. Constraints (\ref{eq:flow_conservation}) ensure flow conservation, and equations (\ref{eq:subtour_elimination}) are subtour elimination constraints represented in cutset form, named Generalized Cut-Set (\textit{GCS}) inequalities. This latter set of constraints ensures that the number of edges with both extremes in \textit{S}, i.e., $|E(S)|$, cannot be greater than the number of vertices in \textit{S} traversed from the \textit{s}--\textit{t} path. This type of constraint is necessary due to the coloring constraints (\ref{eq:coloring_2})--(\ref{eq:coloring_4}), which could generally induce cycles disconnected from the simple path \textit{s}--\textit{t}.
Constraints (\ref{eq:coloring_1}) ensure that each node is colored with at most one color, and constraints (\ref{eq:coloring_2}) ensure that in each shortest path \textit{s}--\textit{t}, there is at least one colored vertex for each color $c\in C$.
The constraints (\ref{eq:coloring_3}) and (\ref{eq:coloring_4}) ensure that a node $i$ can contribute to the $s$--$t$ path with color $c$ only if $i$ is effectively traversed as an intermediate node or as the destination node, respectively. The set of constraints (\ref{eq:coloring_5}) establishes that if a node $i$ contributes to at least one $s$--$t$ path with a specific color $c$, then $i$ must indeed be colored with $c$ in the solution.
Finally, constraints (\ref{eq:domain_1})--(\ref{eq:domain_3}) define the variable domains.

Additionally, a separation procedure is developed for the computationally expensive subtour elimination constraints (\ref{eq:subtour_elimination}).
So, initially, the relaxed problem is considered, meaning the subtour elimination constraints are temporarily omitted. During the resolution process, any violated subtours in the current solution are identified. Regarding the separation routine, a method considered in \cite{ref:sep_procedure} is used, focusing on identifying the strongly connected components in the graph induced by the current solution. Violated \textit{GCS} constraints are dynamically added to the model using a modified version of Tarjan's algorithm (see \cite{ref:gcs_2}), as proposed by \cite{ref:gcs_3}.

\subsection{Meta-heuristic}\label{sec:brkga}
%introduzione BRKGA
The BRKGA is a significant advancement in genetic algorithms, developed to tackle complex and large-scale combinatorial optimization problems. 
It uses a population of solutions represented as vectors of real numbers between 0 and 1, known as random keys. 
A key component in the BRKGA is the decoder, a deterministic function that maps the random-key vectors to the solution space of the specific problem. The decoder ensures that each vector is translated into a solution, maintaining consistency and reproducibility of the results.

In our study, we consider a multi-parent and multi-population BRKGA with bidirectional Permutation-based Implicit Path-Relinking (IPR-Per) (see \cite{ref:brkga_mp_ipr}).
During the evolution process of the considered BRKGA, several key operations are utilized. It starts by creating the first generation of \textit{m} populations and using a seed to generate all the chromosomes. The size of a single population is calculated as $p:= \alpha\cdot n$, where $\alpha\geq 1$ is called population size factor; an elite population is defined as $p_e:=p\cdot pct_e$, where $pct_e\in[0.1, 0.25]$ is the elite percentage parameter; finally, the size of the mutant population is $p_m:=p\cdot pct_m$, where $pct_m\in[0.1, 0.3]$ is the mutant percentage.
In the second step, the decoder converts the chromosomes in the APSPC solutions and consequently computes the fitness values.
If the stopping criteria are not reached, then the next step is to create a new generation and the process is repeated by decoding new populations. In particular, the population of the current generation is divided into two parts according to fitness: the elite population $p_e$ containing the chromosome with the best fitness, and the non-elite population $p_{ne}$ which contains the rest of the chromosomes. The elite individuals are directly copied to the next generation to preserve high-quality solutions. Mutation introduces new random individuals to explore new areas of the solution space. The remaining part of the population, $p(1-pct_e-pct_m)$, is generated by the \textit{multi-parent crossover}. For this crossover, it is necessary to choose three parameters, the number of total parents ($\pi_t$) and elite parents ($\pi_e$) to be selected; the probability that each parent has of passing genes on to their child. The probability is calculated taking into account the bias of the parent, which is defined by a pre-determined, non-increasing weighting bias function ($\phi$) over its rank \textit{r}. Multi-parent crossover allows multiple parents to contribute to the new offspring, increasing genetic diversity. Multi-population evolution enables multiple populations to evolve in parallel and exchange their best individuals, reducing the risk of premature convergence.
Regarding global stopping criteria, we consider two rules. The procedure is interrupted if either the set time limit or the maximum number of consecutive iterations without improvement (\textit{wi}) are reached.

\begin{algorithm}
\caption{decode}\label{alg:decoder}
\begin{algorithmic}[1]
\begin{scriptsize} \State \textbf{input} \textit{chromosome}, \textit{n := number of nodes} (dimension of the chromosome)
\Procedure{decode}{}
    \State Initialize random generator \textit{gen} with seed \textit{chromosome[0]}\label{line:gen}
    \State Reset \textit{nodeColors} to $-1$ for all nodes\label{line:resetColor}
    \For{\textit{i} $\gets 0$ to \textit{n}}
        \State Select a random \textit{color} using \textit{gen}\label{line:randomColor}
        \State \textit{colorCost} $\gets$ \textit{colorCosts[color]}
        \If{\textsc{shouldColorNode}(\textit{i}, \textit{chromosome}, \textit{colorCost}, \textit{gen})}\label{line:shouldColorNode}
        \State \textit{nodeColors[i]} $\gets$ \textit{color}\label{line:colored}
        \EndIf
    \EndFor
    \State \textit{fitness} $\gets$ \textsc{calculateFitness}(\textit{nodeColors})\label{line:fitness}
    \State \Return \textit{fitness}
\EndProcedure
\end{scriptsize}
\end{algorithmic}
\end{algorithm}

Algorithm~\ref{alg:decoder} is designed to transform a chromosome into a solution for the APSPC, evaluating its quality through a fitness function, i.e., it represents the decoder.
The procedure begins with the initialization of a random number generator \textit{gen} using the first value of the chromosome as the seed (line~\ref{line:gen}). This ensures that the random generation operations are reproducible throughout the entire genetic evolution.
In line \ref{line:resetColor}, all nodes are initially uncolored. This is represented by setting \textit{nodeColors} to $-1$ for each node.
The procedure iterates with a \textbf{for} loop over all nodes to determine whether each node should be colored or left uncolored.
In particular, for each node, in line~\ref{line:randomColor} a random color is selected using the random number generator \textit{gen}.
The cost associated with the selected color is calculated by accessing the \textit{colorCosts} vector.
It is then checked whether the node should be colored using the \textit{shouldColorNode} function (line~\ref{line:shouldColorNode}). 
In line~\ref{line:colored}, if the node should be colored, the color is assigned to the node.
Once colors have been assigned to all nodes, the fitness of the solution is calculated using the \textit{calculateFitness} function in line~\ref{line:fitness}, which evaluates the quality of the solution based on the assigned colors.
Finally, the procedure returns the calculated fitness value.

\begin{algorithm}
\caption{shouldColorNode}\label{alg:shouldColorNode}
\begin{algorithmic}[1]
\begin{scriptsize}\Procedure{shouldColorNode}{\textit{node}, \textit{chromosome}, \textit{colorCost}, \textit{gen}}
    \State \textit{nodeDegree} $\gets$ \Call{getNodeDegree}{\textit{node}}
    \State \textit{avgNodeWeight} $\gets$ \Call{getAvgNodeWeight}{\textit{node}}

    \Comment{Phase 1: Probability based on color cost}
    \State \textit{ColorCostFactor} $\gets$ \textit{colorCost} / (\textit{avgNodeWeight} $\cdot$ (\textit{n} - 1))\label{line:ColorCostFactor}
    \If{\textit{ColorCostFactor} $\leq$ 0.1}
        \State \Return \textbf{true}
    \EndIf

    \Comment{Phase 2: Probability based on other node characteristics}
    \If{\textit{chromosome[node]} $\geq$ 0.1}
       \State \textit{NodeProbability} $\gets$ \textit{chromosome[node]} $\cdot$ \textit{nodeDegree} / \textit{avgGraphDegree} $\cdot$ \textit{avgGraphWeight} / \textit{avgNodeWeight}\label{line:nodeProbability_1} 
    \Else \State \textit{NodeProbability} $= 1$\label{line:nodeProbability_2}
    \EndIf
    \State \textit{dis} $\gets$ \Call{UniformRealDistribution}{0.0, 1.0}
    \State \Return (\textit{dis(gen)} $<$ \textit{NodeProbability})
\EndProcedure
\end{scriptsize}
\end{algorithmic}
\end{algorithm}
   
The next function to be analyzed is \textit{shouldColorNode}. Algorithm~\ref{alg:shouldColorNode} is designed to determine whether a node in the graph should be colored based on the node's characteristics. The procedure begins by getting the degree of the input node and the average weight of the edges incident to the node (\textit{avgNodeWeight}).
The decision process is divided into two phases to ensure a balanced evaluation, it is sufficient that one of the two phases is verified for the node to be colored.
In Phase 1, the procedure calculates the \textit{ColorCostFactor} as a function of the color cost and \textit{avgNodeWeight} (line~\ref{line:ColorCostFactor}). This probability assesses the cost-effectiveness of coloring the node. If the ratio is very low, the node is colored with certainty. Intuitively, this means that we color the node if the cost of coloring is relatively small compared to the benefit we gain from coloring it.
Phase 2 focuses on other characteristics of the node. The procedure calculates the \textit{NodeProbability} as a function of the ratio between \textit{nodeDegree} and \textit{avgNodeWeight}, and the chromosome gene associated with the node (line~\ref{line:nodeProbability_1}). This operation allows us to determine how important it is to color a node based on the number of connections and the strength of those connections (average edge weight). If these values indicate that the node is influential in the network, then the probability of coloring it increases. If the gene is too low, the probability is set to one to avoid invalidating the probability calculation.
Finally, a random number is generated using a uniform distribution between 0.0 and 1.0, and the node is colored if this random number is less than \textit{NodeProbability} (line~\ref{line:nodeProbability_2}). The procedure returns the boolean result, indicating whether the node should be colored or not.

The \textit{calculateFitness} function evaluates the fitness of a solution by calculating the aggregate path cost between all pairs of nodes within the graph, based on their color assignments. Initially, it computes an overall color cost derived from color assignments. Subsequently, the algorithm iterates over each node pair to determine the shortest path between them, applying a modified Dijkstra algorithm that incorporates the color constraints. If a valid path exists, its cost is added to the aggregate path cost. If no valid path is found, the algorithm designates the solution as infeasible and halts further calculations.

\subsection{Strong Learner Splitting and \#colors Selection}
The number of colors (i.e., the different WLs that compose the entire SL) available to color the nodes of a given graph is chosen using the function defined below, denoted as $cd~:~\mathbb{R}~\rightarrow~2~\mathbb{Z}~+~1$. Given a real number \textit{x}, this function returns the largest odd integer less than \textit{x} or returns 3 if the largest integer less than \textit{x} is 2. Formally:

$$cd(x):=\begin{cases}
    3 & \text{if } \lfloor x\rfloor = 2\\
    \lfloor x\rfloor -1 & \text{if } \lfloor x\rfloor\in 2\mathbb{Z}\setminus\{2\}\\
    \lfloor x\rfloor & \text{if } \lfloor x\rfloor\in 2\mathbb{Z}+1.
\end{cases}$$

The exact number of colors, \textit{\#colors}, available for the graph $G=(V,E)$ is given by evaluating the function \textit{cd} in the average number of nodes present in all classical shortest paths, i.e., without the coloring constraint.

\begin{equation}\label{eq:cd_function}
\#colors = cd\Big(\dfrac{2}{|V|\cdot (|V|-1)}\sum_{(i,j)\in E | i < j} d(i,j)\Big),
\end{equation}

where $d(i,j)$ is the number of nodes present in the classical shortest path between \textit{i} and \textit{j} calculated using the Dijkstra algorithm.

\section{Performance Evaluation}\label{sec:perf}
%misure
This section illustrates an in-depth performance evaluation campaign conducted to assess the benefits of the proposal in terms of both optimization and network-relevant aspects. Two experimental campaigns will be described in order to accomplish this task: \textit{(i)} Model Evaluation Campaigns; and \textit{(ii)} Network Evaluation Campaigns.

\subsection{Model Evaluation Campaigns}\label{sec:model_evaluation}
In this section, we summarize the results of our computational experiments on the meta-heuristic defined. In particular, we conduct an in-depth analysis of the impact of various network characteristics on the effectiveness and efficiency of the entire defined system. 

The BRKGA has been implemented in C++ using clang version 14.0.3. For the compilation, the C++17 standard was set using the CMAKE$\_$CXX$\_$STANDARD 17 specification in the CMake configuration file.
All the optimization computational tests were conducted using an Apple M2 Max processor with CPU 12‐core and GPU 38‐core and 96 GB LPDDR5 of RAM running macOS Ventura 13.3.
\\

\subsubsection{Instances and Parameter Setting}\label{sec:instances}
In order to evaluate the performance of the proposed approach, a set of instances was generated as described below.
The set is composed of random topology networks, each of which is identified by a unique combination of the following parameters: number of nodes ($n$), edge density ($d$), and color cost ranges ($cr$). In particular, we considered: four values for the number of the nodes, i.e., $n\in\{10, 15, 25, 30\}$; four values for the edge density, that determines the number of the edges $\#e = d\cdot n(n-1)/2$, with $d\in\{0.25, 0.35, 0.45, 0.55\}$; and four ranges of values for the color cost, i.e., $cr_1 = [1,125]$, $cr_2 = [50,150]$, $cr_3 = [75,175]$, $cr_4 = [100,200]$. For each instance, the number of colors is uniquely determined by the function (\ref{eq:cd_function}). More in detail, given a certain number of nodes, we start by generating the minimum spanning tree $G = (V, E)$ first to ensure connectivity, then we randomly add edges to $E$, until the needed number of edges, determined by the edge density parameter, is reached. The costs of the edges are determined as a sample from a uniform distribution in the interval $[1, 200]$. The color costs are determined as a sample from a uniform distribution in the color costs value range parameter.
For each scenario, identified by a given combination of values of $n$, $d$, and $cr$, we generated six different random instances, for a total of 384 instances, by varying the seed used to initialize the random number generator. We organized each set into four classes, based on edge density, named $\{\text{ED}_i\}_{i=1}^4$.

For the metaheuristic parameters, we carried out a preliminary tuning phase using \textit{irace}, a tool that performs an automatic configuration to optimize parameter values (refer to \cite{ref:irace} for details). This tuning was done using four random instances of each of the $AD_i$ sets. 
Table~\ref{tab:parameters} summarizes the tuned parameters of the BRKGA, grouping them into three sets: \textit{Operator}, \textit{IPR-Per} and \textit{Others}.

\begin{table}[ht]
\caption{Tuned BRKGA parameters.}\label{tab:parameters}
\centering
{\begin{tabular}{llllllllll}
\hline\noalign{\smallskip}
\multicolumn{5}{l}{$Operator$} & \multicolumn{3}{l}{$IPR-Per$} & \multicolumn{2}{l}{$Other$} \\[-6pt]
\multicolumn{5}{l}{\hrulefill} & \multicolumn{3}{l}{\hrulefill} & \multicolumn{2}{l}{\hrulefill}\\
$pct_e$ & $pct_m$ & $\pi_t$ & $\pi_e$ & $\phi$ & $sel$ & $md$ & $pct_p$ & $\alpha$ & $m$ \\
0.1 & 0.6 & 3 & 1 & $1/r^2$ & randS & 0.15 & 0.85 & 20 & 2 \\
\noalign{\smallskip}\hline
\end{tabular}}
\end{table}

\subsubsection{Experimental Results}\label{sec:results}
The summary table will be presented by grouping instances according to their density class $\text{ED}_i$ and the number of nodes. Each row in the tables refers to a subset of instances from a given set that share the same edge density and, where specified, the same number of nodes. These are indicated by the descriptor in the \textit{Set} column, where the acronym “ED" stands for edge density and “N" stands for nodes. Furthermore, all time values are measured in seconds. Table~\ref{tab:brkga_results} provides detailed information on the results obtained by applying BRKGA to the set of all instances. Each row reports the average values for the following parameters: the number of available colors in the instances ($\#colors$), calculated using the \textit{cd} function; the time taken by the metaheuristic to identify the obtained solution ($BestTime$ (s)); the total execution time (\textit{Time} (s)); the number of deployed nodes ($\#NDy$); the total solution cost (\textit{Cost}); color-related costs ($Cost_c$); and path cost ($Cost_p$). The number of referred instances is 24 for each row aggregating on both the edge density and the number of nodes, and 96 for the $AVG$ rows aggregating only on the edge density. For all the experiments, we set the time limit equal to 900 seconds and the maximum of consecutive iterations without improvement \textit{wi} to 10. 

%commenti sui risultati - BestTime e Time
Analyzing the behavior of the average best time, it increases as expected as both the number of nodes and the density increase. However, the effect of the number of nodes is more significant compared to the density, while still remaining below 1 minute. In particular, as shown in Table~\ref{tab:brkga_results}, we observe that with 10 nodes, the \textit{BestTime} consistently stays within the 0.08--0.32 second range, regardless of density. With 15 nodes, it increases significantly compared to 10 nodes, but remains manageable, ranging between 1.22 and 2.70 seconds. With 25 nodes, there is an increase, but still limited, in fact, it rises to 16.99 seconds for \textit{ED1} and 21.62 seconds for \textit{ED3}. With 30 nodes, the highest recorded \textit{BestTime} is observed, with values ranging from 34.21 seconds for \textit{ED1} to 55.65 seconds for \textit{ED4}. In general, it is observed that as the density increases, the \textit{BestTime} increases linearly for each number of nodes. This increase becomes greater as the number of nodes increases.
In addition, for each density class, it is noted that as the number of nodes increases, the \textit{BestTime} increases in a non-linear manner.
Similarly, the total runtime of the BRKGA follows a linear trend as the density increases for each number of nodes %(as shown in Fig.~\ref{fig:plot_density}).(b),
and a non-linear trend as the network size increases for each density class. 

%commenti sui risultati - #colors
Regarding the average number of colors identified by the \textit{cd} function, it is observed that, on average, the number of colors increases as the density decreases. Specifically, in all instances with 10 and 15 nodes, $\#colors$ is always equal to the minimum available, which is 3. With 25 nodes, the average ranges from 3.08 in the \textit{ED3} class to 3.17 in \textit{ED1}, while in the \textit{ED4} class, all instances have $\#colors$ equal to 3. Overall, 5 instances with 5 colors were recorded. With 30 nodes, the highest $\#colors$ values are recorded, ranging from 3.08 in the \textit{ED4} class to 3.42 in \textit{ED1}. In total, 4 instances with 7 colors and 3 instances with 5 colors were recorded. Therefore, for each density class, as the number of nodes increases, $\#colors$ also increases. These trends can be explained by the fact that, in fully random topologies with a greater number of nodes and/or relatively low density, it is more likely to find, on average, the shortest path with a higher length, which requires the use of more colors, as expected from the definition of the \textit{cd} function.

%commenti sui risultati - nodi deploiati
As expected, $\#NDy$ increases with the total number of nodes in the network. For example, in the case of 10 nodes and density class \textit{ED1}, the average number of deployed nodes is 4.21, while with 30 nodes in the same class, it increases to 14.79. This trend is consistent across all classes, confirming that as the graph size increases, more nodes are involved in the deployment of learning models and VNFs necessary to ensure network security coverage. With the same number of nodes, it is observed that as density increases, the number of deployed nodes tends to increase. For instance, for $N=15$, $\#NDy$ increases from 6.92 in density class \textit{ED1} to 9.00 in class \textit{ED2}, and 8.08 in class \textit{ED4}.  
The scalability of the proposed model is evident from the way it adapts to networks of varying sizes and densities. The increase in $\#NDy$ with the growth in both the number of nodes and density shows that the model can handle larger and more complex network topologies. This scalability is crucial for next-generation networks, where the number of nodes and connections will continuously increase, requiring an efficient distribution of learning functions across the network.

%commenti sui risultati - costi
The increase in the number of nodes has a significant impact on the total costs for each density class. For example, observing the results in the table, for 10 nodes and \textit{ED1}, the \textit{Cost} is around $2\cdot 10^4$, while for 30 nodes in the same density class, the cost rises to approximately $6.5\cdot 10^4$. This increase is attributable to the rise in both deployment costs ($Cost_c$) and shortest path costs ($Cost_p$), as larger networks require the distribution of VNFs across more nodes and covering longer distances. Density, however, follows a different trend. As density increases, $Cost_p$ decreases because the paths between nodes become shorter. Nevertheless, $Cost_c$ tends to rise slightly with the increase in density, as more nodes are needed to manage the more connected network. Therefore, since $Cost_p$ constitutes the vast majority of the total cost for each set of instances (over 90\%), the average total cost decreases, as can be seen from the AVG rows.

\begin{table}[ht!]\caption{Detailed results of the BRKGA}\label{tab:brkga_results}
\scriptsize\centering
\resizebox{!}{2.6cm}{
\begin{tabular}{cccccccc}
\hline
Set   & $\#colors$  & \textit{BestTime} & \textit{Time} & $\#NDy$ & Cost & $Cost_c$ & $Cost_p$ \\ 
\hline
N10ED1 & 3.00  & 0.08  & 0.16     & 4.21         & 20605.7      & 516.5     & 20089.2  \\ 
N15ED1 & 3.00  & 1.22  & 2.82     & 6.92         & 30081.7      & 941.7     & 29140.1  \\ 
N25ED1 & 3.17  & 16.99 & 40.80    & 12.38        & 50684.4      & 1641.1    & 49043.2  \\ 
N30ED1 & 3.42  & 34.21 & 88.02    & 14.79        & 64995.9      & 1920.0    & 63075.9  \\ 
\hline
AVG   & 3.15 & 13.13 & 32.95    & 9.57         & 41591.9      & 1254.8    & 40337.1  \\ 
\hline
N10ED2 & 3.00 & 0.17  & 0.45     & 4.58         & 11526.4      & 579.4     & 10947.0  \\ 
N15ED2 & 3.00 & 1.40  & 5.29     & 9.00         & 21894.2      & 1200.5    & 20693.6  \\ 
N25ED2 & 3.17 & 19.04 & 57.06    & 14.50        & 36370.6      & 1908.2    & 34462.4  \\ 
N30ED2 & 3.25 & 37.27 & 110.52   & 16.83        & 47228.7      & 2190.7    & 45037.9  \\ 
\hline
AVG   & 3.10 & 14.47 & 43.33    & 11.23        & 29254.9      & 1469.7    & 27785.2  \\ 
\hline
N10ED3 & 3.00 & 0.32  & 0.81     & 4.88         & 9140.1       & 618.0     & 8522.1   \\ 
N15ED3 & 3.00 & 2.70  & 7.61     & 7.42         & 16526.6      & 865.9     & 15660.1  \\ 
N25ED3 & 3.08 & 21.62 & 59.56    & 14.21        & 28602.1      & 1900.3    & 26701.7  \\ 
N30ED3 & 3.17 & 36.21 & 163.72   & 17.75        & 38783.9      & 2277.9    & 36505.9  \\ 
\hline
AVG   & 3.06 & 15.21 & 57.92    & 11.06        & 23263.2      & 1415.5    & 21847.6  \\ 
\hline
N10ED4 & 3.00 & 0.30  & 1.15     & 4.96         & 8436.8       & 681.6     & 7755.2   \\ 
N15ED4 & 3.00 & 2.09  & 8.26     & 8.08         & 13872.0      & 1076.8    & 12795.2  \\ 
N25ED4 & 3.00 & 16.94 & 92.46    & 16.79        & 25438.2      & 2081.6    & 23356.5  \\ 
N30ED4 & 3.08 & 55.65 & 185.47   & 17.25        & 30632.8      & 2004.6    & 28628.2  \\ 
\hline
AVG   & 3.02 & 18.75 & 71.84    & 11.77        & 19594.9      & 1461.2    & 18133.8  \\ 
\hline
\end{tabular}}
\end{table}

\subsection{Network Evaluation Campaigns}   

During a further experimental campaign, we compared the performance of the data plane devices when dealing with an entire ML model and when, instead, the model is decomposed following our deployment approach. We measured the time to obtain the classification outcome -- namely \textit{classification time} --  and the \textit{throughput} guaranteed by the networking devices that execute the additional and AI-related task. In addition, to evaluate the detouring imposed on the shortest path nature of the network due to the coloring constraint, we introduced the \textit{
AWDelay} metric. Further detail about this metric will be given in Section~\ref{sec:awdelay}

\begin{figure}[!ht]
\centering
\includegraphics[width=3in]{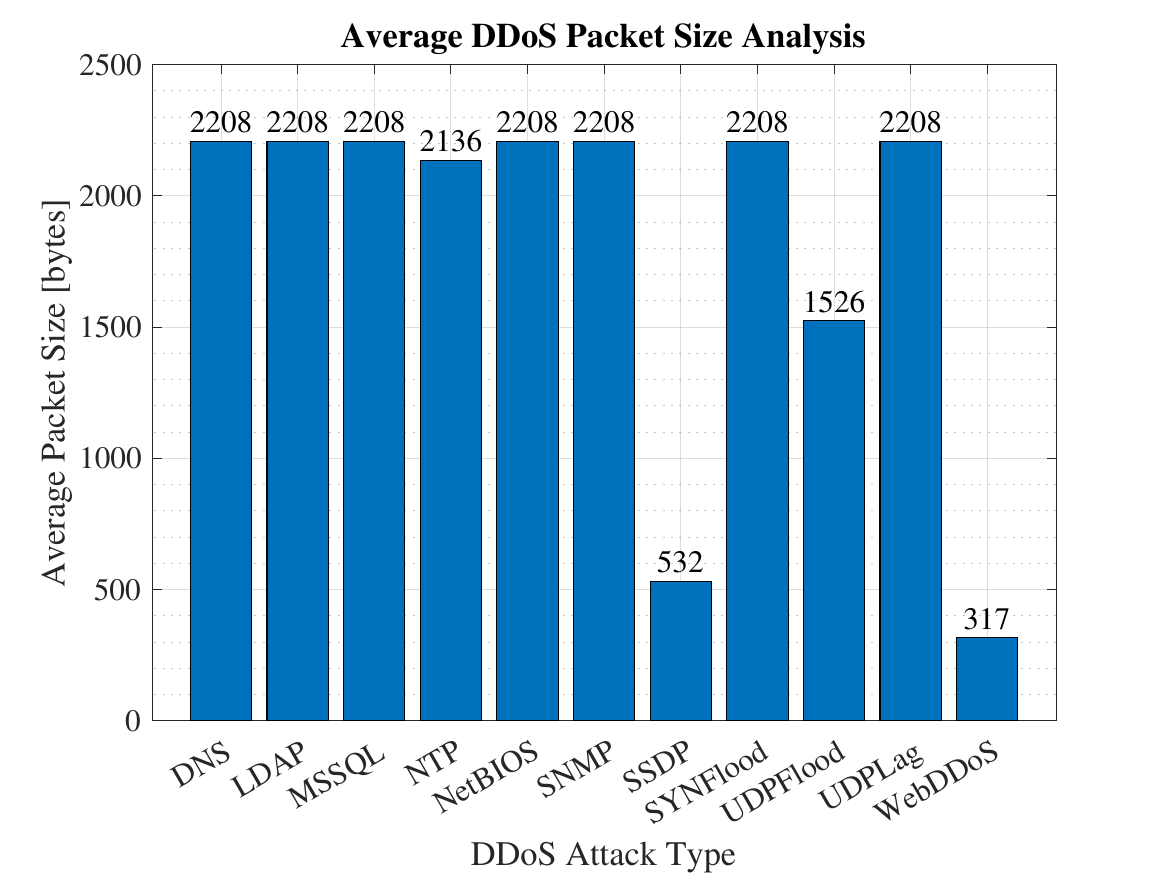}
\caption{Average Packet Size for DDoS attack in CIC-DDoS2019.}
\label{fig:ddos_packet_size}
\end{figure}

The objective is to assess that under heavy network load, e.g., volumetric Distributed Denial of Service (DDoS), the reduced workload imposed on the single data plane device will lead the network to scale well in these critical situations guaranteeing the forwarding activities. %In the wake of these considerations, 
We tested the network by considering different attack intensities, starting with 100 pkt/s generated by each of the attackers and reaching 1000pkt/s with an incremental step of 100 pkt/s.  
To characterize the size of the DoS/DDoS packets, we analyzed the DDoS evaluation dataset (CIC-DDoS2019) \cite{cicddos}. The dataset contains real-world data, recorded by the Canadian Institute for Cybersecurity (CIC), representing the most common DDoS attack types -- characterized by means of 80 network features -- such as SYN flooding, UPD DDoS, DNS-based DDoS, WebDDoS, and many others. On the basis of the analysis conducted on the average packet size (\textit{Avg Packet Size} feature), we uniformly chose the attack packet size in the range [317,2208] bytes (see Fig.~\ref{fig:ddos_packet_size}). %The data 
Following the work in \cite{Doshi2021Jan}, in order to parameterize the attack scenario with respect to the network topology, we considered a number of attackers that is set to 50\% of the total hosts of the network. 

In order to recreate a real experimental scenario, we generated typical benign background traffic based on the CIC-IDS 2018 \cite{CIC}. In particular, we considered the dataset days Wednesday-14-02-2018\_TrafficForML\_CICFlowMeter, Wednesday-21-02-2018\_TrafficForML\_CICFlowMeter, Wednesday-28-02-2018\_TrafficForML\_CICFlowMeter. We analyzed the probability distribution of the interarrival times registered in the benign flows (more than 1.5 million samples) finding an exponential distribution with a $\lambda=0.4$. To generate the benign background traffic we used the Distributed Internet Traffic Generator (D-ITG) generator \cite{BottaDP12, waqas} and set the lambda equal to 0.4; while the packet size is uniformly distributed within a range of [16, 360] bytes. 

Finally, since our proposal extends the shortest-path nature of the networks for the sake of security, we evaluated how much the traditional short path is affected by the security and cooperative behavior constraints. In other words, we evaluated the \textit{traffic detouring} from the traditional shortest path that is caused by complying with network security constraints. The network topologies used to test the experiments are publicly available at \cite{ourGithub}. 

We evaluated the proposal scalability under three topologies of increasing dimensions: the first one with 10 nodes and 25 edges, for which the value of $\#colors$  computed by Eq.\ref{eq:cd_function} is $3$ (i.e., SL is splitted into three WLs); the second one with 25 nodes and 48 edges and a computed $\#colors$ equal to $5$ ; and the third, bigger, topology with 30 nodes and 51 edges, for which $\#colors= 7$. 
For the sake of the experimental campaign, we chose as an SL a Random Forest (RF) and, therefore, the WLs are represented by the Decision Trees (DTs) that compose the RF. This choice does not affect the general applicability of the proposed Split-AI technique. Any type of ensemble learning model can be employed, with the only constraint being the instruction set supported by the target programmable network element to encode the model within it. The performance of each considered SL are summarized in Tab.\ref{tab:SL_performance}. As previously mentioned, $\#colors$ correspond to the number of WLs in the SL. 

\begin{table}[ht]\caption{SL Performance Analysis}\label{tab:SL_performance}
\centering
\begin{tabular}{ccccc}
\hline
    & \textit{Accuracy}   & \textit{FPR}   & \textit{FNR}   & \textit{F1-Score} \\ \hline
\textit{SL3}   & $\sim$98\% & 1.4\% & 0.15\%  & $\sim$97\% \\ 
\textit{SL5}   & $\sim$99\% & 1.3\% & 0.5\%  & $\sim$98\% \\ 
\textit{SL7}   & $\sim$99\%  & 0.2\% & 0.3\%  & $\sim$99\% \\ \hline

\end{tabular}
\end{table}

The 15-node topology previously considered is not used for these experiments as the calculated value of $\#colors$ was found to be identical to that of the 10-node topology, and thus it adds little to the experiment.

The chosen topologies allow to test the scalability degree of the proposal while increasing the model complexity and therefore the amount of WLs that need to be deployed to obey and guarantee the network security coverage. According to \cite{lee2020switchtree}, these are appropriate SL complexities when dealing with network traffic classifications. However, the proposal is general enough to be extended to more complex models, making it adaptable for other AI-relevant tasks. 

The proposal has been implemented using P4-enabled virtual PDP, namely BMv2 \cite{bmv2} that are based on the v1Model architecture. Due to the limited instruction set of the P4 language (it does not support basic operations such as division, exponentiation or logarithm), we extracted 43 features of the CIC-IDS 2018. The P4 code that implements the models and the associated feature extractor will be publicly made available.\footnote{The code will be made available on GitHub repository at \cite{ourGithub}.}
\\
\subsubsection{Evaluating Shortest Path Detouring: AWDelay}\label{sec:awdelay}
Given a pair of source and target nodes $(s,t)$, we denote with $SP(s,t)$ the cost of the classical shortest path between \textit{s} and \textit{t}, and equivalently, we denote with $SP_C(s,t)$ the cost of the shortest path obtained for the problem with coloring constraints.

We can define a weighted average of the delays as a function of the lengths of the classical shortest paths. Let $delay(s,t)$ be the relative delay between the constrained shortest path and the classical one between source \textit{s} and target \textit{t}, i.e.,

$$delay(s,t):=\dfrac{SP_C(s,t)-SP(s,t)}{SP(s,t)},$$

then the weighted average of delays is defined as follows:

$$AWDelay:=\dfrac{2}{|V|\cdot(|V|-1)}\cdot \sum_{(i,j)\in E |i<j}\overline{w_{ij}}\cdot delay(i,j),$$

where $\overline{w_{ij}}$ is the normalization of the following weights that depend on the length of the classical shortest paths defined as:
\begin{center}$w_{ij}:= e^{length(SP(i,j))}.$\end{center}

%\begin{comment}

\begin{figure*}
    \centering
    \subfloat[]
    {
    \label{fig:classification_time.1}
    \includegraphics[width=0.65\columnwidth]{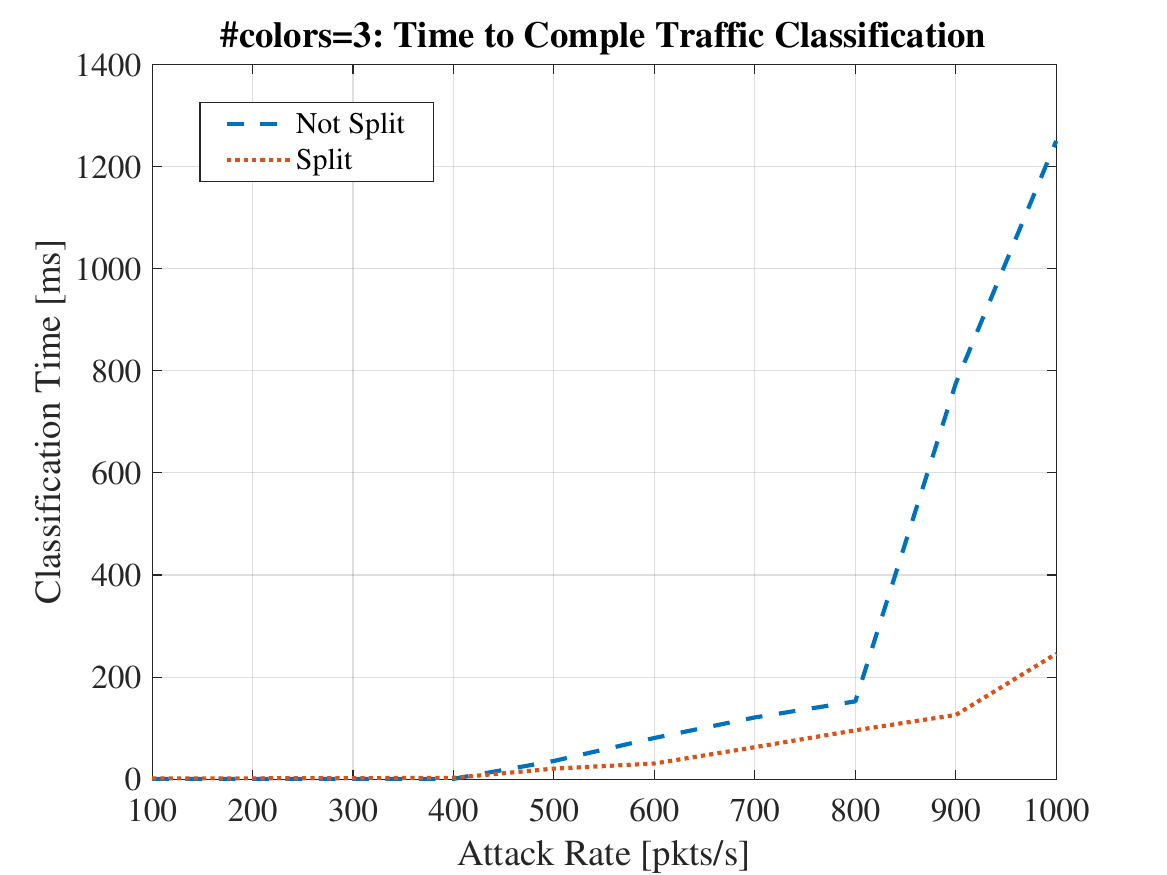}
    }
    \subfloat[]
    {
    \label{fig:classification_time.2}
    \includegraphics[width=0.65\columnwidth]{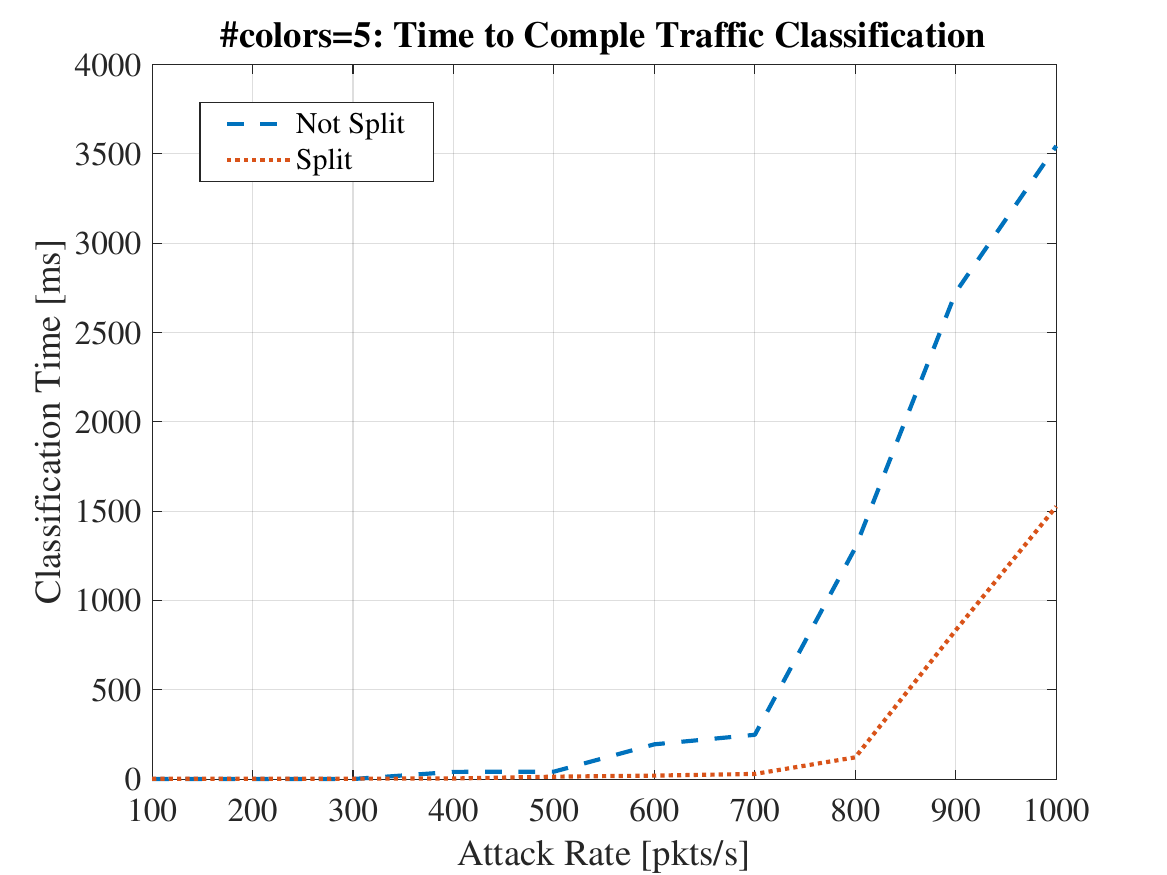}
    }
    \subfloat[]
    {
    \label{fig:classification_time.3}
    \includegraphics[width=0.65\columnwidth]{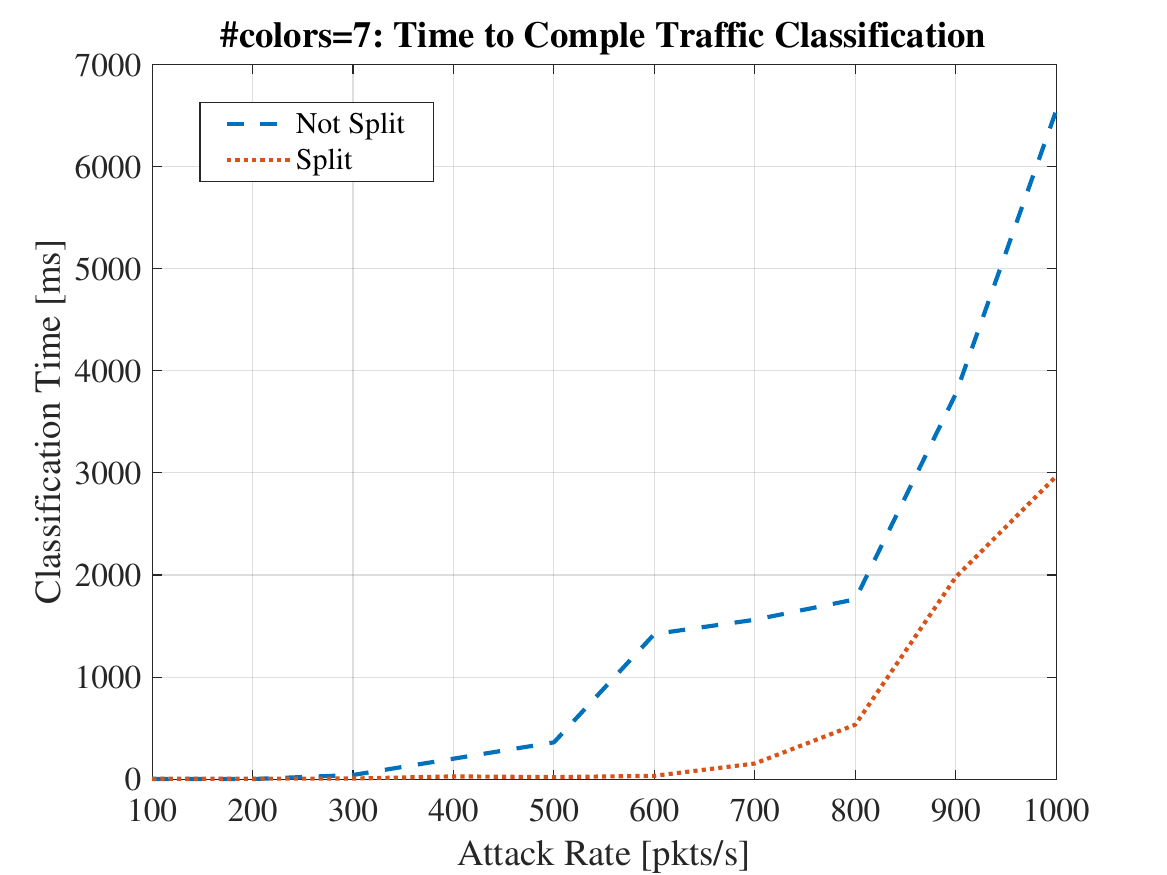}
    }

    \caption{Average Classification Time for Experimental Scenarios: a)$\#colors=3$, b)$\#colors=5$ ,$\#colors=7$.}
    \label{fig:classification_time}
\end{figure*}

\begin{figure*}
    \centering
    \subfloat[]
    {
    \label{fig:throughput.1}
    \includegraphics[width=0.65\columnwidth]{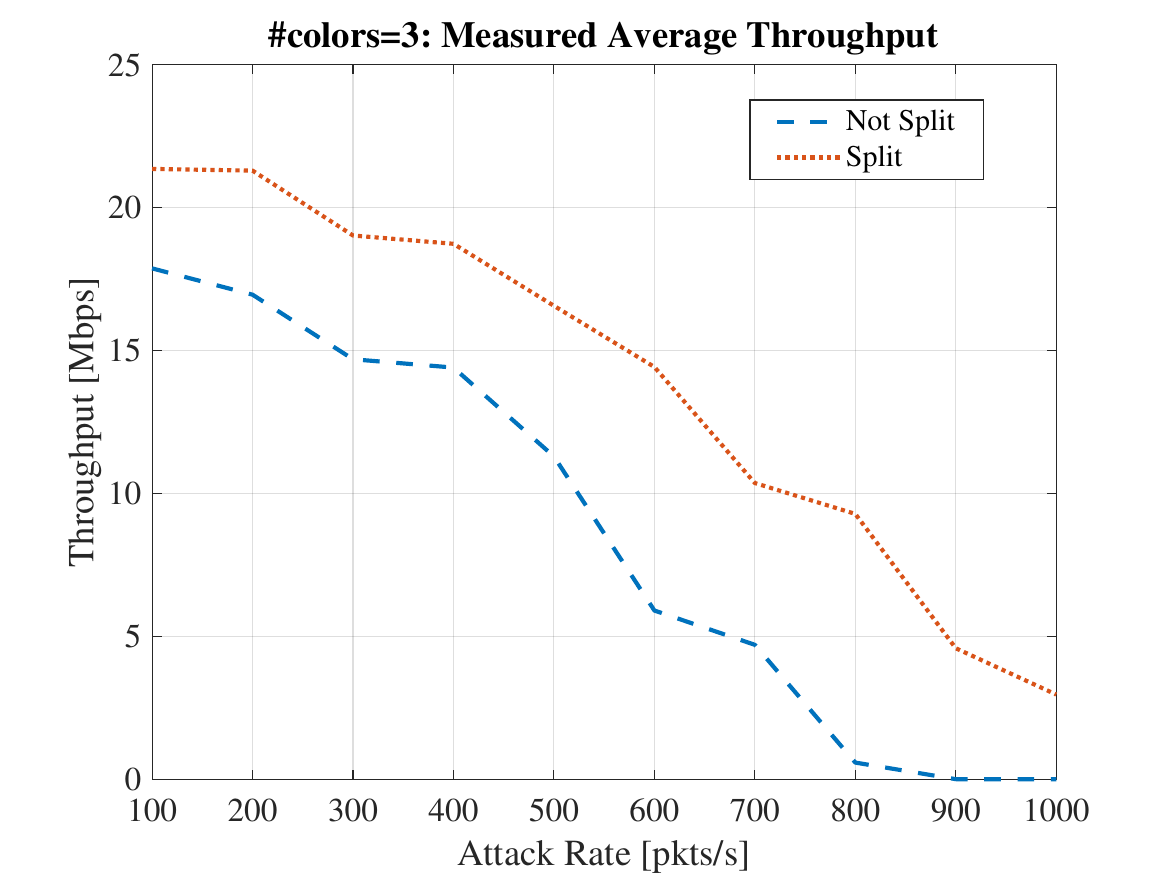}
    }
    \subfloat[]
    {
    \label{fig:throughput.2}
    \includegraphics[width=0.65\columnwidth]{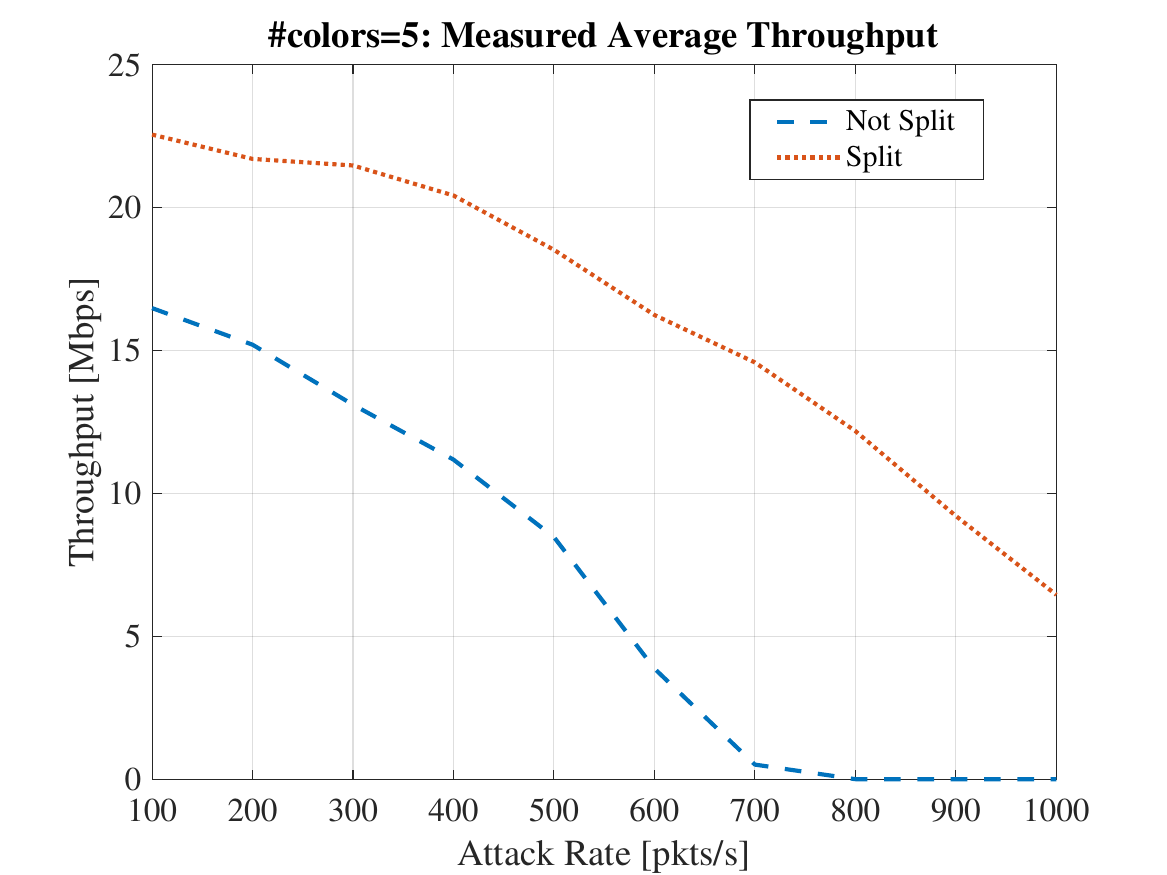}
    }
    \subfloat[]
    {
    \label{fig:throughput.3}
    \includegraphics[width=0.65\columnwidth]{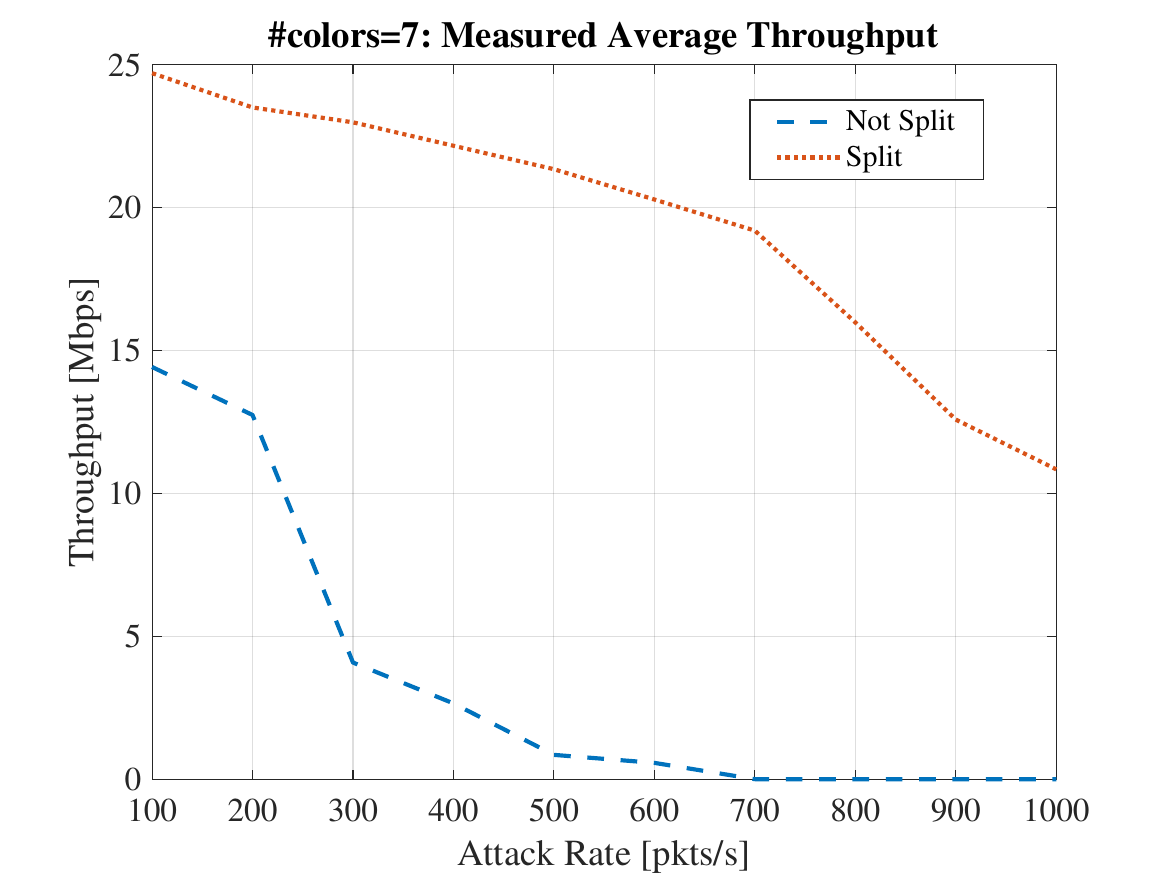}
    }
    \caption{Average Throughput for Experimental Scenarios: a)$\#colors=3$, b)$\#colors=5$ ,$\#colors=7$. }
    \label{fig:throughput}
\end{figure*}
%\end{comment}

\subsubsection{Classification Time Analysis}

We also analyzed the average classification time of the networking devices within the proposed distributed approach under increasing traffic loads. 

In Fig.~\ref{fig:classification_time.1}, the achievable average classification time under a varying attack rate is shown. 
With the first small topology (10 switches, 50 hosts of which 25 are attackers) -- which requires a SL composed of three WLs to guarantee the security coverage -- it can be observed that while the amount of handed packets is around 200--400 pkts/s the SL-VNF configuration performs better, showing an average classification time that is about 60\% less than the WL-VNF (an average of 0.62 ms of the SL-VNF against 1.7 ms of the WL-VNF). This is due to the additional intermediate communication that happens between the PDPs to get the final classification. However, as the attack rate intensifies and the switches become overwhelmed with network packets to analyze, this advantage diminishes, allowing the WL-VNFs configuration to demonstrate its strengths in handling critical attack situations. The differences can be appreciated when the attack rate is in the range of 600--800 pkts/s, with the classification time more than halved. Under heavy attack load, 900--1000 pkts/s, the SL-VNF configuration is not able to timely handle the classification tasks, reaching a maximum time to complete classification which is more than 1000 ms against the $\sim$ 200 ms achieved through the adoption of the proposed model splitting and distribution paradigm. 

In Fig.~\ref{fig:classification_time.2} the results with the medium network topology (25 network switches and 125 hosts – 75 attackers) and a SL composed of five WL-VNF. In this case, due to the lesser model complexity, the benefits of the proposal can be appreciated starting from 300--400 pkts/s and it shows its effectiveness around 500--600 pkts/s by reducing the time to complete the classification of more than 90\%. Even under the highest attack rate (1000 pkts/s), the reduction achieved by the proposal is more than 50\% ($\sim$1500 ms with the proposal against $\sim$3600 ms with the SL-VNF configuration). 

This trend is confirmed by the experiments carried out with the largest topology (see Fig.~\ref{fig:classification_time.3}), in which the optimization problem suggested an SL with seven WLs to cope with network security coverage. In this case, the highest complexity of the SL-VNF leads the network to be unable to timely handle classification tasks starting from an attack rate of 300 pkts/s. At 500 pkts/s the gap starts to be prominent, with an average classification time of $\sim$360 ms for the SL-VNF against 20 ms for the split configuration. When the attack rate is around 1000 pkts/s, the benefits of the proposal are indeed highlighted allowing the network to adapt to the huge attack rate, showing a reduction of 55\% in the average classification time. 

In light of the considerations made so far, it can be concluded that as the size of the network topology and the load it is subjected to increase, using a split-AI approach to distribute the workload within programmable data planes, allows for an effective integration of complex AI-relevant tasks within the network, but also a scalable and adaptable solution to network changes. These results shed light on the importance of split-AI approaches to cope with the upcoming seamless and tight integration of networking and AI, for future 6G networks.
\\

\subsubsection{Throughput Analysis}

In a further test campaign the average throughput of the PDPs in both configurations, i.e., SL-VNF and WL-VNF, is measured by varying the network topologies and the related value of $\#colors$. This is to demonstrate that the proposed approach of optimizing the distribution of active IDS features is scalable in terms of network devices' capacity in managing network traffic.

It is observed in Figs.~\ref{fig:throughput} that the WL-VNFs deployment setting shows the best gain for the network, both in terms of throughput and delays, with the increase in the amount of traffic generated by the distributed malicious hosts. When considering the SL-VNF configuration, the throughput experienced by the network devices decreases as the SL complexity increases (from three to seven WLs), mainly due to the increasing number of WLs that need to be queried on a single PDP. With the simplest SL, the average network throughput starts to drop below 5 Mbps when the attack rate is 700 pkt/s, quickly approaching 0 Mbps at 800 pkt/s. This trend worsens when considering more complex SLs. The $\#colors=5$ scenario shows that the network throughput drops to zero when approaching an attack rate of 600--700 pkt/s. Even worse is the case of the most complex SL ($\#colors=7$), whose overhead causes the average network throughput to approach zero starting from an attack rate in the range of 400-500 pkt/s. In such situations, data plane devices experience substantial degradation in their forwarding capabilities.

However, when the SL is split and distributed across the network, the computational load imposed on the PDP devices is alleviated, making it possible to consider the integration of even complex AI models within the network without affecting the normal network operation too much. In fact, when considering the $\#colors=3$ scenario and the split configuration, the average network throughput starts to drop below 5 Mpbs with an attack rate of 900--1000 pkt/s.
Considering an attack rate in the range of 100--500 pkt/s, we saw a 20\% increase in throughput on average. With a higher attack rate, this advantage improved further ($\sim$50--55\%), up to the point where the advantages of the distributed approach ensure that the network is still able to guarantee a minimum throughput while with the SL-VNF the network is completely down again.
The advantages of the proposed approach become more evident as the complexity of the SL increases and the size of the network expands. When it is necessary to split a SL into five WLs to cover the network, the resulting reduction of the computational burden in each device preserves even more the average network throughput. Indeed, the average network throughput is in the range [$\sim$6, $\sim$15] Mbps even under attack rates of 700--1000 pkts/s, where instead the non-split configuration causes the average throughput measured on the PDPs to be zero.
Finally, in the $\#colors=7$ network topology, the complexity of the SL causes significant performance degradation starting from attack rates of 400 packets per second (leading to a rapid zeroing of the average throughput), while the proposed distributed approach improves scalability. This method effectively manages the computational overhead, allowing the network to handle large attack volumes while maintaining a satisfactory level of throughput.

Nonetheless, a truly zero-cost solution does not exist yet. The execution of models still imposes a measurable impact on network throughput, with an observed average value of approximately 35 Mbps when no SL/WL-VNFs are active within the switch.
This limitation stems from the technological constraints of current networking devices which are not yet inherently designed to fully support the seamless integration of networking and AI workflows.
However, it is expected that these issues will be resolved in future 6G networks, which will likely incorporate advanced, high-performance chips capable of significantly increasing computational power. 
Having said that, the advantages of the proposed distributed and split AI approach are clear, making it a viable solution for supporting AI-relevant tasks within current as well as future PDP devices. Finally, it is important to highlight a key feature of the proposed approach: it can effectively operate (without any modification) with both encrypted and unencrypted network traffic, as it relies exclusively on header information, which is always transmitted in plaintext.
\\

\subsubsection{Shortest Path Detouring Analysis}\label{sec:detouring_analysis}
To evaluate the impact of the coloring constraints on network performance, we also conducted an analysis of the \textit{AWDelay} metric introduced previously. Specifically, we assessed the impact of network density and size on path detours by analyzing the average weighted delay.

Table~\ref{tab:AWDelay} presents the average \textit{AWDelay} values grouped by the number of nodes $N$ and the density class $ED$. The \textit{AWDelay} values shown for each combination of $N$ and $ED$ represent the average computed across all instances discussed in Section~\ref{sec:results}.
The AVG row reports the average calculated based on the nodes, while the column AVG shows the average relative to the density. Additionally, the row labeled VAR indicates the variance of all \textit{AWDelay} values for each number of nodes $N$, providing a measure of data dispersion and allowing us to assess the variability with the number of nodes.

\begin{figure*}
\centering
\subfloat[][\emph{N10}]
{\includegraphics[width=.25\textwidth]{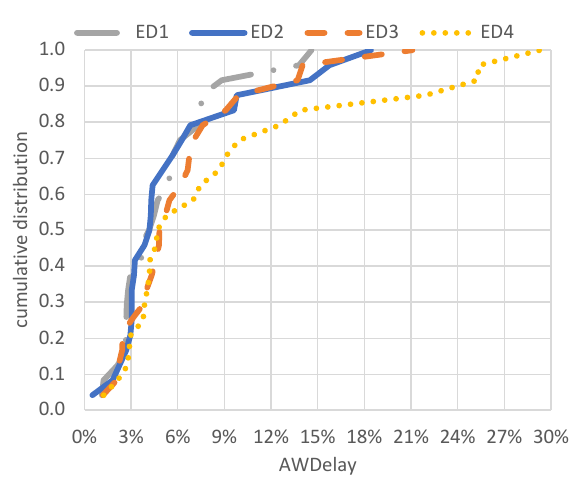}} \subfloat[][\emph{N15}]
{\includegraphics[width=.25\textwidth]{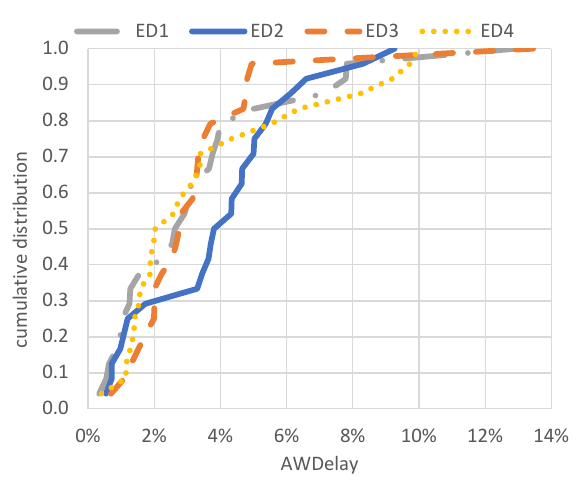}} \subfloat[][\emph{N25}]
{\includegraphics[width=.25\textwidth]{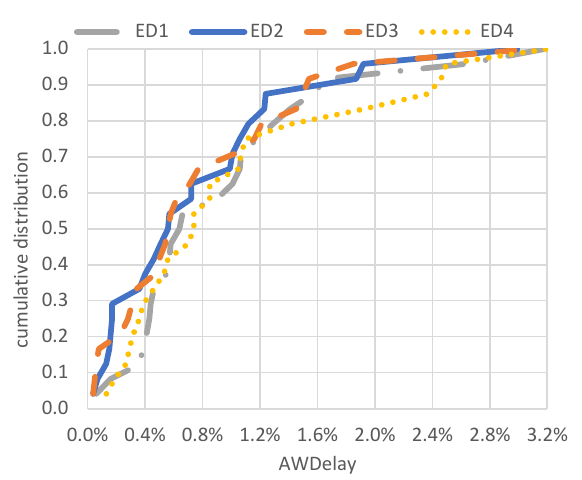}}
\subfloat[][\emph{N30}]
{\includegraphics[width=.25\textwidth]{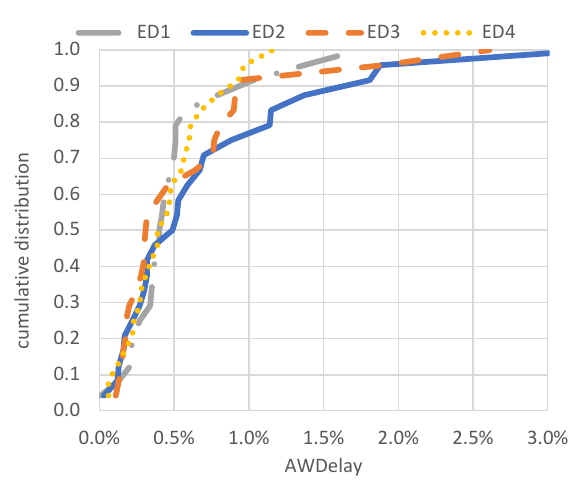}}
\caption{Cumulative distribution of \textit{AWDelay} for the density classes for each node class.} \label{fig:cumulative_distribution}
\end{figure*}    

The plots shown in Fig.~\ref{fig:cumulative_distribution} represent the cumulative distribution of \textit{AWDelay} for the density classes for each node class. Thus, each curve shows the cumulative percentage of recorded results that exhibit an \textit{AWDelay} less than or equal to a specific value indicated on the x-axis.

%N10
For $N = 10$, a clear upward trend in the curves is observed, where a high percentage of observed values (around 60\%) is concentrated within the lower \textit{AWDelay} range (0--5.5\%), especially for the first three density classes. On the other hand, the results for \textit{ED4} show generally higher delays, but more spread out over a wider interval. Specifically, the curves associated with the first three density classes show a rapid accumulation around 5\% \textit{AWDelay}, while the \textit{ED4} curve shows a slower accumulation, suggesting a more dispersed distribution of delays, with the presence of paths experiencing higher delays. In this class of nodes, the minimum and maximum \textit{AWDelay} values are 0.52\% and 29.30\%, respectively, with an overall average of 6.58\%.

%N15
For $N = 15$, the graph in Fig.~\ref{fig:cumulative_distribution}.(b) shows a behavior similar to what was previously observed, but with some significant differences. First of all, for all density classes, 80\% of delays are below about 5\%. A slight difference is seen in the \textit{ED3} class, where about 95\% of the values are concentrated in the lower \textit{AWDelay} range (0--5\%). Another difference is that in this class, the trends of the four curves are quite similar. The minimum and maximum \textit{AWDelay} values are 0.34\% and 13.45\%, respectively, with an overall average of 3.54\%.

%N25
Compared to the previous plots, the graph with 25 nodes (Fig.~\ref{fig:cumulative_distribution}.(c)) shows a more concentrated distribution of \textit{AWDelay} values. All the plots reach 90\% of the cumulative distribution at lower \textit{AWDelay} values compared to the previous plots. This indicates that most of the paths in networks with 25 nodes experience lower delays, concentrating below around 2.5\% \textit{AWDelay}. Specifically, the curves for the first three density classes show almost identical behavior, with very rapid accumulation (90\%) for delays below about 1.5\%. The \textit{ED4} curve shows a similar trend, although it has a slightly more gradual increase, suggesting greater variability in delays compared to the other density classes, but still well-contained compared to cases with fewer nodes. The minimum and maximum \textit{AWDelay} values are 0.04\% and 3.19\%, respectively, with an overall average of 0.88\%.

%N30
Similarly, in the plots of Fig.~\ref{fig:cumulative_distribution}.(d), as previously observed for the instances with 25 nodes, the \textit{AWDelay} values are concentrated within a very narrow range (up to 3.5\%). Similar to the previous case, all the curves reach 90\% of the cumulative distribution at \textit{AWDelay} values below about 2\%. Specifically, the curves representing \textit{ED1}, \textit{ED3}, and \textit{ED4} show almost identical behavior, with a high percentage of observed values (90\%) having delays below about 1\%. The \textit{ED2} curve shows a similar trend but with a slightly more gradual increase, indicating greater variability in delays compared to the other density classes. The minimum and maximum \textit{AWDelay} values are 0.01\% and 3.30\%, respectively, with an overall average of 0.57\%.
The curves associated with 30 and 25 nodes converge much more quickly compared to those for 10 and 15 nodes. This suggests that, as the number of nodes increases, the effect of network density becomes less pronounced, leading to more similar delay distributions.

%Size
The results of the experiments, as shown in Table~\ref{tab:AWDelay}, indicate that the average weighted delay behaves consistently as the network grows in size. Specifically, \textit{AWDelay} significantly decreases with an increasing number of nodes. For example, in networks with 10 nodes in the density class $ED1$, the average delay reaches around 5\%, while for networks with 30 nodes, the delay drops to approximately 0.5\%. This trend can also be observed in the average delay, which decreases from 6.58\% with 10 nodes to 0.57\% with 30 nodes. This indicates that the overhead introduced by the coloring constraints becomes less significant in larger networks, making the approach more scalable and efficient as the network grows.

%Density
Interestingly, when varying the density for a fixed number of nodes, except for the case with 10 nodes, the average \textit{AWDelay} remains almost constant. The variance of all \textit{AWDelay} values decreases from $3\cdot 10^{-3}$ for $N = 10$ to $3\cdot 10^{-5}$ for $N = 30$. This behavior is attributed to the fact that as density increases, and consequently, the number of available paths increases, the probability of significant detours from the classic shortest path decreases, thus mitigating any further delay reduction. For example, networks with $N = 30$ and higher density classes (such as $ED4$) consistently show lower \textit{AWDelay} values, supporting the hypothesis that denser networks provide more direct alternative paths even with coloring constraints. The stability of \textit{AWDelay} across different density classes reinforces the robustness of our approach, as the method maintains a consistent balance between security and efficiency without significantly compromising network performance, even in denser topologies.

\begin{figure}[!ht]
\centering
\includegraphics[width=3in]{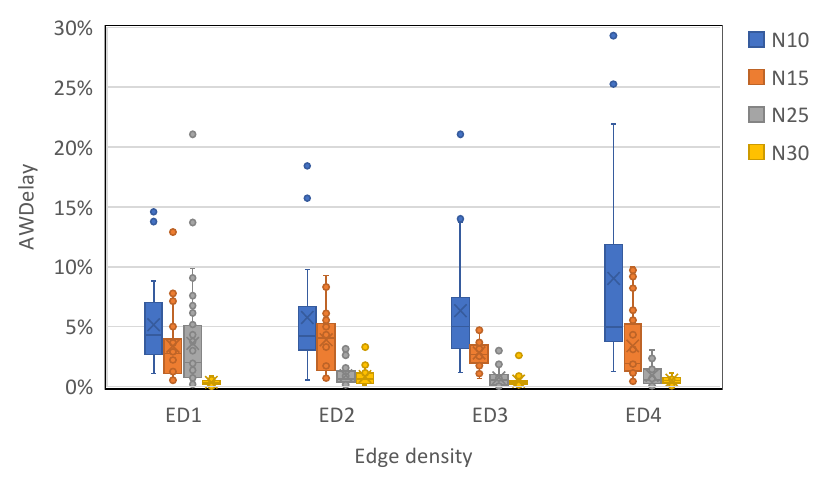}
\caption{Box plots of \textit{AWDelay} for each edge density class.}
\label{fig:box_plot_density}
\end{figure}

This trend is further supported by the variability observed in Fig.~\ref{fig:box_plot_density}, where the box plots illustrate the distribution of \textit{AWDelay} across different densities. In particular, the interquartile ranges expand in sparser networks, showing greater variability in path efficiency due to the limited number of feasible paths that meet the coloring constraints. The box plots also highlight that in more connected networks, such as those with $ED4$, the \textit{AWDelay} distribution is more compact, suggesting a more uniform detour behavior.

\begin{table}[ht]\caption{\textit{AWDelay} results for density and number of nodes}\label{tab:AWDelay}
\centering
\begin{tabular}{ccccc|c}
\hline
    & \textit{N10}   & \textit{N15}   & \textit{N25}   & \textit{N30}   & AVG   \\ \hline
\textit{ED1}   & 5.16\% & 3.37\% & 0.94\% & 0.50\% & 2.50\% \\ 
\textit{ED2}   & 5.77\% & 3.92\% & 0.78\% & 0.73\% & 2.80\% \\ 
\textit{ED3}   & 6.34\% & 3.26\% & 0.79\% & 0.58\% & 2.74\% \\ 
\textit{ED4}   & 9.04\% & 3.60\% & 1.02\% & 0.46\% & 3.53\% \\ \hline
AVG  & 6.58\% & 3.54\% & 0.88\% & 0.57\% &         \\ \hline
VAR  & 0.003 &	0.001 &	0.0001 &	0.00003   &     \\ \hline
\end{tabular}
\end{table}

\section{Conclusion and Future Works}\label{sec:concl}

In this paper, we explored the benefits of the INC paradigm and the programmable nature of networks combined with distributed AI and split-AI techniques with the aim of improving the security of upcoming 6G networks. We considered a split-AI approach through which complex ensemble (SL) models are broken into lightweight functional blocks to be executed on PDPs as a chain of VNFs (WL-VNFs). The goal of these functions is to detect malicious behaviors that may occur in the network. We formulated an optimization problem, All-Pairs Shortest Path Coloring, that intelligently distributes the WL-VNF components on PDPs while taking into account both the shortest path nature of the network and the constraint of reconstructing the decomposed SL by concatenating the distributed WL-VNFs that compose it. To efficiently solve the APSPC problem, we also designed a meta-heuristic approach. The results demonstrate that the joint combination of INC and distributed AI not only overcomes the limitations of implementing complex AI models on PDP devices but also significantly increases the scalability and preserves the forwarding capabilities of AI-enhanced PDPs, especially under heavy traffic attack conditions.

\bibliographystyle{IEEEtran}
\bibliography{biblio}

\begin{IEEEbiographynophoto}{Mattia Giovanni Spina} is a PhD student at the University of Calabria (Italy). His research interest is in the area of security in future generation networks and distributed AI in-network architectures.
\end{IEEEbiographynophoto}
\vspace{-1.0 cm} 

\begin{IEEEbiographynophoto}{Floriano De Rango} 
is full professor of Telecommunications at the University of Calabria (Italy). His research interests include security in wireless and IoT networks and networking solutions for V2X systems.
\end{IEEEbiographynophoto}
\vspace{-1.0 cm} 

\begin{IEEEbiographynophoto}{Antonio Iera} 
is full professor of Telecommunications at the University of Calabria (Italy). His research interests include next generation mobile and wireless networks and the Internet of Things. He is currently Editor in Chief of the Elsevier Computer Networks journal.
\end{IEEEbiographynophoto}
\vspace{-1.0 cm}

\begin{IEEEbiographynophoto}{Edoardo Scalzo}
is a junior researcher of Operations Research at the University of Calabria (Italy). His research interests include network optimization, logistics and combinatorial optimization.
\end{IEEEbiographynophoto}
\vspace{-1.0 cm}

\begin{IEEEbiographynophoto}{Francesca Guerriero}
is a full professor of Operations Research at the University of Calabria, Italy. Her primary research interests revolve around network optimization, logistics, combinatorial optimization, and the intersection of optimization and big data.
\end{IEEEbiographynophoto}
%\vspace{-1.5 cm}

\vfill

\end{document}